# Flow Characterization in Triply-Periodic-Minimal-Surface (TPMS) based Porous Geometries: Part 1 - Hydrodynamics

Surendra Singh Rathore[a], Balkrishna Mehta[a*], Pradeep Kumar[b], Mohammad Asfer[c]

[a] Department of Mechanical Engineering, Indian Institute of Technology Bhilai, Raipur, 492015, (Chhattisgarh) India

[b] Numerical Experiment Laboratory (Radiation & Fluid Flow Physics), Indian Institute of Technology Mandi, Mandi- 175075, (Himachal Pradesh), India

[c] Department of Mechanical Engineering, College of Engineering, Dawadmi, Shaqra University, Shaqra, 11911, Saudi Arabia

## Abstract

The modeling of flow and heat transfer in porous media systems have always been a challenge and, the extended Darcy transport models for flow and equilibrium and non-equilibrium energy models for heat transfer are being used for macro-level analysis, however, the limitations of these models are subjected to porous geometry. The forced convective flow of an incompressible viscous fluid through a channel filled with four different types of porous geometries constructed using the Triply-Periodic-Minimal-Surface (or TPMS) model, are presented in this study. Four TPMS lattice shapes namely; Diamond, I-WP, Primitive, and Gyroid are created with identical porosity, and three different types of porous media are further generated for each porous geometry to investigate the relationship of shape-tortuosity, microporosity, and pore size on permeability and inertial drag factors. A pore-scale direct numerical simulation approach is performed for the first two types of porous media by solving the Navier-Stokes equations. The specific microporosity is quantitatively induced in the solid region where Darcy-Forchheimer-Brinkman model is solved, whereas the Navier-Stokes equations is solved for the fluid region in the third type of porous media. The results reveal that the validity of Darcy flow regime is very narrow up to $Re \sim 4$ for the Primitive lattice (Type 1) while for Diamond lattice (Type 2), it extends up to $Re \sim 20$. For $Re > 20$, Darcy regime is not valid for any lattice types. For lower porosity (Type 1, $\varepsilon = 0.32$) the inertial drag is found to be minimum in I-WP lattice and maximum in Gyroid lattice while, for higher porosity ( Type 2, $\varepsilon \sim 1$), Primitive lattice has minimum and I-WP lattice has maximum value of inertial drag, respectively.

# 1. Introduction

The interconnected network of pores of various shapes and sizes characterizes porous structures in general [1]. When fluid travels through this porous network, it encounters varied levels of stagnation, separation, recirculation, and attachment within the pores, resulting in a superior mixing or dispersion effect in the porous medium [2–5]. The increased heat and mass transport in porous media is extremely useful in mini-to-micro scale channels due to the tortuous nature of the path and a very high surface area to volume ratio [6, 7]. In these mini-scaled systems, turbulent flow regimes are difficult to achieve, and the flow is largely in the laminar regime, nevertheless, the utilization of porous media in these systems is advantageous [8, 9]. These flow systems can be found in a variety of applications, including electronic device heat management, nuclear reactors, solar receivers, and avionics, to mention a few.

One of the popular thermal management devices, Loop Heat Pipe (LHP) as shown in the Fig. 1 is composed primarily of three components: an evaporator, a condenser, and an adiabatic section (liquid and vapour lines, where no heat transfer occurs), out of which, evaporator is the most crucial component to design, build, and control operation [10–12]. The porous medium used in the evaporator fulfills two objectives; firstly, to provide a sufficient capillary pressure difference for condensed liquid to flow back to the evaporator, and secondly, to provide augmented heat and mass transfer due to higher surface area to volume ratio, and dispersion effect. The conjugate heat transfer, liquid-vapor phase change, multiphase flow, capillarity and wetting, and porous media flow are all key transport processes occur in the evaporator [13]. In such cases, monitoring and controlling the complex transport processes in evaporator (porous media) becomes more significant and therefore requires the rigorous analysis of the transport mechanism.

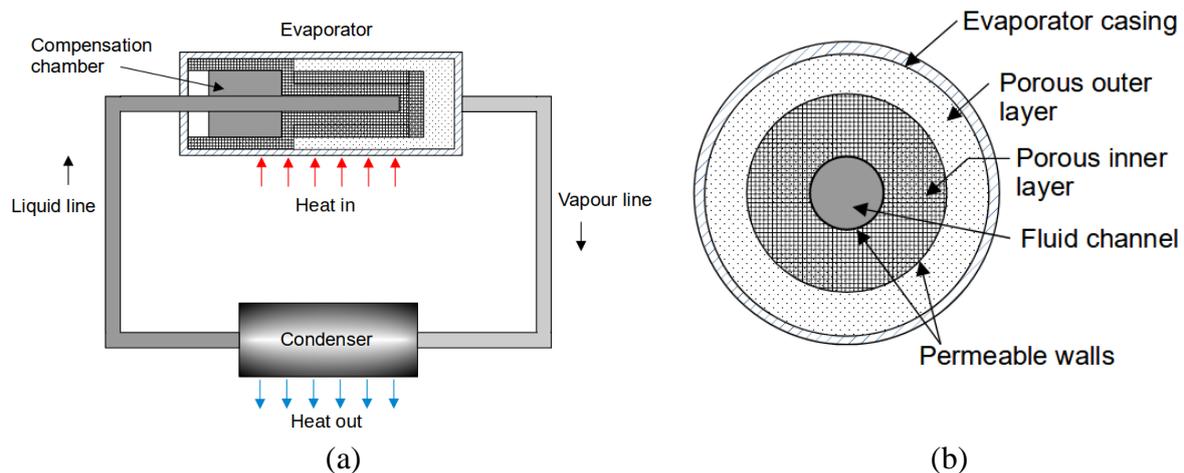

**Figure 1: (a) Schematic of a Loop Heat Pipe, and (b) Cross-section view of evaporator.**

Generally, three basic types of porous structures are utilized in heat pipes: grooved, wick, and sintered. The grooves are highly ordered patterns which are inexpensive to manufacture but functionally poor for capillary action and heat transfer. The sintered porous structures are the most expensive to manufacture due to their high stochasticity, however have the best heat transfer and capillary action capabilities. Moreover, wick structures are optimum choice for intermediate performance and manufacturing [14, 15].

For the numerical analysis of transport phenomena in the porous media, there are two approaches which are popularly being followed: direct and indirect modelling. In the indirect modelling, actual porous geometry is not constructed, instead, the fluid domain is virtually assigned as the porous region using associated viscous and inertial resistances to it and then the porous media flow simulations (popularly known as Darcy advection models[4]) are performed [16–18]. This approach is sufficiently effective in hydrodynamic studies of single-phase flow though porous media, and based on the flow rate regimes or porosity values, different models are being used (refer to Eq. A1 to A3 in the Appendix) [19–21]. However, in case of multiphase flow in porous media, this approach may not be as much effective to handle the complicated relationship of relative permeability and capillary pressure with the saturation [22–24]. On the other hand, in the direct modelling, the porous structure (or solid matrix) is digitally constructed as a CAD model and the pore-scale simulation is performed over this geometry [23, 25, 26]. Both single phase and multiphase flow can be effectively and accurately handled due to lesser involvement of models. Furthermore, on considering the direct modelling approach, two kinds of solid matrix are found, ordered and random. Geometrical reconstruction of random porous structures is a challenging task in and of itself, and it also necessitates very expensive equipment such as X-ray CT scanners and MRI machines, making the entire process uneconomical [27–31]. Processing scanned tomographic images into a 3D digital model is similarly time-consuming and error-prone, resulting in a geometric representation that differs from the original porous structures. On the other hand, with the advent of 3D printing ordered solid matrix is preferred due to controlled performance of the system and repeatability; thus, many researchers have attempted to create simplified ordered structures (both two and three dimensional) to resemble with the porous media. In some cases, mostly circular or square shaped obstructions were used in both patterned and random distributions [32, 33]. Whereas for other cases, spherical shaped particles were used in Simple cubic (SC), face centered cubic (FCC), body centered cubic (BCC) and hexagonal closed packing (HCP) arrangements [34–36]. The errors associated with the geometry will therefore may hamper the system's performance and may not provide the repeatability. As a result, when compared to random structures, using ordered porous structures for analysis and application is both cost-effective, accurate, and performance consistent [37, 38].

The pore-scale numerical simulations in the ordered porous structures have been performed for the flow in order to extract the parameters for the macro-scale modelling and also to determine the limit for the applicability of the Darcy law. TPMS based ordered porous geometries are used and numerical analyses are performed in the present work to characterize the quantities such as permeability, inertial drag factor, and also the flow range up to which the Darcian regime prevails, are investigated. The manuscript is organized as follows: section 2 defines the problem statement, followed by section 3 which contains the governing equations and boundary conditions for pore-scale simulations. The solution methodology is described in Section 4, the results and discussion are presented in Section 5, and lastly, the present work's conclusions are presented in Section 6. The appendix describes the equations, some relationships, and parameters for the macro-scale modeling of the flow in porous medium.

## 2. Problem statement

In this study, the hydrodynamics of the flow within a porous medium of square-cross section has been characterized by using the pore-scale simulations. The ordered porous structures are generated using Triply-Periodic-Minimal-Surfaces (TPMS) method, and four different TPMS lattices, namely; Diamond, I-WP, Primitive, and Gyroid are developed as shown in Fig. 2. Further, three scenarios of solid matrix of porous medium are considered as, solid matrix as solid zone; solid matrix as fluid zone; and solid matrix as porous zone as shown in Fig. 3. The lattices created are described below:

### 2.1 TPMS lattice and their level-set equations

A minimal surface is a surface with a mean curvature of zero is known as a TPMS and is infinite and repetitive in three dimensions. To develop a minimal surface, different mathematical approaches were proposed, such as, nodal surface approximation and phase field approach. The level-set approximation strategy, on the other hand, is the simplest and most widely used method and has been used in the present work [39, 40].

Level-set equations are a group of trigonometric functions in three dimensions that satisfy the equality condition $f(X,Y,Z) = C$. Four sets of equations as given below, and can produce four different lattices [41, 42].

1. *Schwarz-Diamond (Diamond)*
$$f(X,Y,Z) = \cos X \cos Y \cos Z - \sin X \sin Y \sin Z \qquad (1)$$

2. *Schoen-IWP (I-WP)*
$$f(X,Y,Z) = 2(\cos X \cos Y + \cos Y \cos Z + \cos Z \cos X) \\ - (\cos 2X + \cos 2Y + \cos 2Z) \qquad (2)$$

3. *Schwarz-Primitive (Primitive)*
$$f(X,Y,Z) = \cos X + \cos Y + \cos Z \qquad (3)$$

4. *Schoen-Gyroid (Gyroid)*
$$f(X,Y,Z) = \sin X \cos Y + \sin Y \cos Z + \sin Z \cos X \qquad (4)$$

Where, $X = 2a\pi x, Y = 2b\pi y, Z = 2c\pi z$ and *a, b, c* are constants related to the unit cell size in corresponding x, y, z directions. Iso-surfaces divide space into equal-subvolumes when the level-set equation is assessed at *C = 0*. The constant can be used to manage these subvolumes, allowing the total volumes to be enlarged or contracted by offsetting from zero in either the normal or opposite directions. By treating one of the volumes separated by the minimal surface as the solid region and the other as the void (pore) region, a TPMS lattice material based on these zero-thickness surfaces can be constructed. The volume enclosed by the minimal surfaces such that $f(X,Y,Z) > C$ or $f(X,Y,Z) < C$ is designated as solid matrix. Using above equations, an STL file containing information of surfaces is generated by open-source software MS-Lattice [40] which uses the surface minimization algorithm in MATLAB environment. The lattices which are used in the present study are shown in Fig. 2.

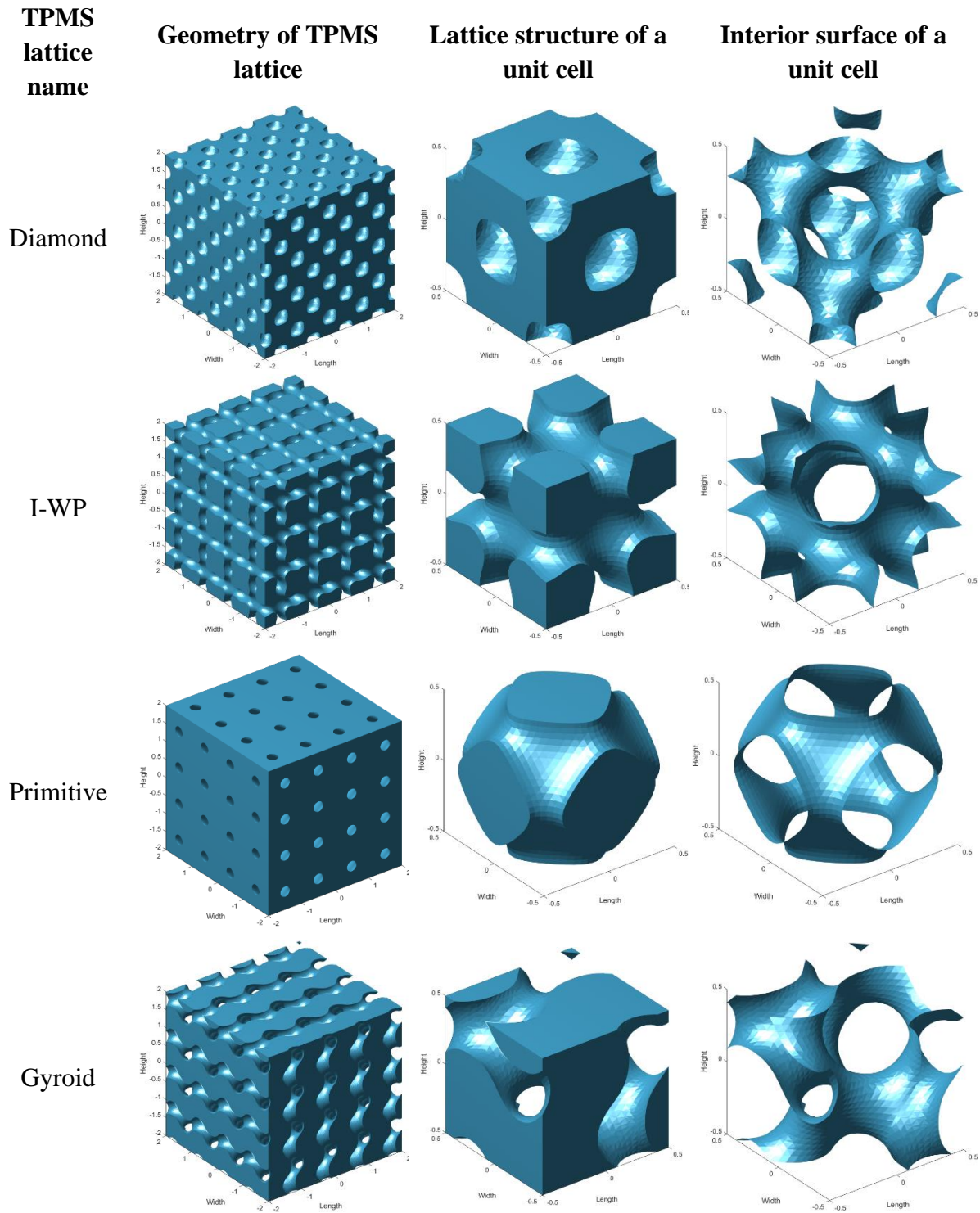

**Figure 2: Solid region and interior surface of four porous structures used in the study based on TPMS lattices.**

Furthermore, the three scenarios for each lattice by treating the solid region as solid, fluid, or porous regions are considered. The effective porosity of the porous media may differ by different treatments of the solid region. The details of different scenarios of each lattice are shown in the Fig. 3.

| Description | Type 1 | Type 2 | Type 3 |
|---|---|---|---|
| Treatment of solid region in the solver | Solid zone (No equation) | Fluid zone (Navier-Stokes equation) | Porous zone (Darcy-Forchheimer-Brinkman equation) |
| Physical representation of the porous medium created | 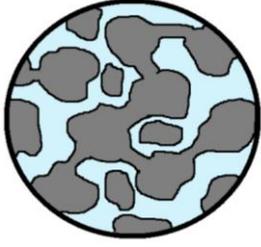 | 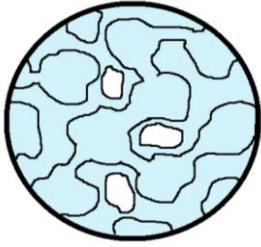 | 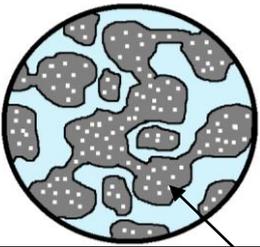 Micro porous zone |
| Effective porosity calculation ($\varepsilon_0$ is the original porosity or void fraction) | $\varepsilon = \dfrac{V_f}{V_f + V_s}$ <br> $\varepsilon = \dfrac{1}{1 + \left(\dfrac{V_s}{V_f}\right)}$ <br> $\varepsilon = \varepsilon_0$ <br> $\varepsilon_{eff} = 0.32$ | $\varepsilon = \dfrac{V_f}{V_f + V_s}$ <br> $\varepsilon = \dfrac{1}{1 + \left(\dfrac{V_s}{V_f}\right)}$ <br> $\varepsilon \to 1$ <br> $\varepsilon_{eff} \sim 1$ | $\varepsilon = \dfrac{V_f + \varepsilon^* V_s}{V_f + V_s}$ <br> $\varepsilon = \dfrac{V_f + \varepsilon^*\left(V_f/\varepsilon_0 - V_f\right)}{V_f + V_s}$ <br> $\varepsilon = \varepsilon_0 + \varepsilon^*(1-\varepsilon_0)$ <br> $\varepsilon_{eff} = 0.54$ |

**Figure 3: Description and effective porosity calculation in three different types of scenarios used for the solid region.**

The effective porosity in Type 1 is same as original, which is equal to the voidness fraction (0.32) of the lattice. In case of Type 2, effective porosity is approximately equal to 1 as voids are only separated by zero thickness wall and different from normal channel flow as zero thickness wall will not allow to grow boundary layers on the impermeable wall of the channel, however they offer resistances to flow. By the induction of microporosity ($\varepsilon^* = 0.32$) in Type 3 lattice, the effective porosity increases to 0.54. The pore-scale simulation has been performed on the pores of the lattice for first two lattices and porous media modelling (Darcy-Forchheimer-Brinkman model) has been used in the microporous region, and, flow parameters in the microporous zone have been obtained from experimental results of Hetsroni (2006) and are shown in Table A3[43]. Therefore, by different treatments of solid zone in the TPMS lattice, three different level of porosity is obtained as low porosity to super-porosity and a total of 12 scenarios of flow in porous media have been considered in the present work.

Usually, three different kinds of Reynolds number definitions have been in use in the literature, as (a) based on Brinkman screening length ($Re_K = \rho U K^{1/2}/\mu$), (b) based on pore size ($Re_p = \rho U d_p/\mu$), and (c) based on channel size ($Re = \rho U L/\mu$) and keeping the flow scenario in the loop heat pipe, the current work is performed for the Reynolds number based on channel size and the mass flow rate (which are related as $\dot{m} = \mu L Re$), for three sets of porosity values $\varepsilon_{eff} = 0.32, 0.54, and \sim 1$. Darcy-Forchheimer model (Appendix A) is used as the reference model to calculate the flow properties of porous media, namely permeability

(or its dimensionless form Darcy number) and inertial drag factor through estimation of pressure developed across the lattice.

## 3. Governing equations

### 3.1 Governing equations for flow at pore-scale

At pore scale, the 3D Navier-Stokes equations are solved for above lattices with the assumptions of steady, incompressible, and laminar flow. The fluid is Newtonian with constant thermo-physical properties ($\rho = 998.2 \ kg/m^3$ and $\mu = 0.001003 \ kg/m \cdot sec$). Due to small size of the pores through which the fluid is being transported, the gravitational force is neglected. The equations which are solved for the fluid zones are written as follows:

$$\nabla \cdot \vec{v} = 0 \tag{5}$$

$$\rho(\vec{v} \cdot \nabla)\vec{v} = -\nabla p + \nabla \cdot \left[ \mu \left( \nabla \vec{v} + (\nabla \vec{v})^T \right) \right]; \quad for \ \vec{v} = u\hat{i} + v\hat{j} + w\hat{k} \tag{6}$$

Where, the $\vec{v}$ and $p$ are the pore-scale velocity and pressure, respectively. $\rho$ and $\mu$ are respectively, the fluid density and dynamic viscosity.

### 3.2 Governing equations for flow in microporosity

The flow in microporous region (in Type 3) is modelled by using Darcy-Forchheimer model, as given below:

$$\nabla \cdot \vec{V} = 0 \tag{7}$$

$$\nabla P = -\frac{\mu}{K}\vec{V} - \frac{\rho C_F}{\sqrt{K}}\vec{V}|\vec{V}|; \quad for \ \vec{V} = U\hat{i} + V\hat{j} + W\hat{k} \tag{8}$$

Where $p$ and $\vec{V}$ are Eularian extrinsic phase averaged pressure and velocity. The pressure gradient consists of two terms: linear and quadratic. The linear term is dominant in the low flow rates, called Darcian regime, whereas the quadratic term is dominant in the relatively higher flow rates, called as non-Darcian regime.

### 3.3 Boundary conditions

Fig. 4 shows the representation of TPMS lattice with a translational periodic boundary condition between inlet and outlet to resemble with a channel of infinite length or hydrodynamically developed flow condition[44]. The other boundaries i.e., top, bottom, front and back are impermeable wall of the channel. The level-set wall which separates two zones (solid & void) are internal walls and no-slip boundary conditions are used on them. These walls are treated as permeable walls (or interior) for microporosity case (Type 3). The mathematical representation of boundary conditions is summarized in Table 1.

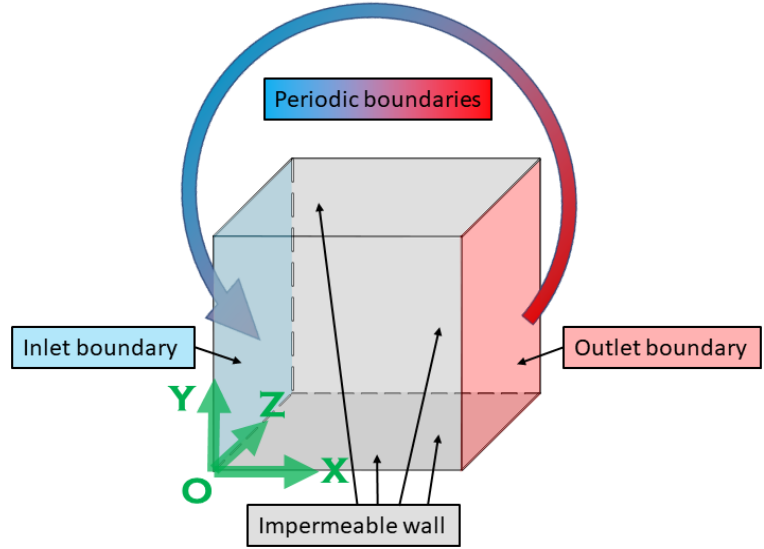

**Figure 4: Representation of TPMS lattice with the translational periodic condition in one direction.**

**Table 1: Details of the boundary conditions for pore-scale simulations.**

| Name of the surface | | Boundary condition | Mathematical description |
|---|---|---|---|
| Inlet-Outlet | | Translational periodic | $u(x_0) = u(x_0 + L); \quad \frac{\partial p}{\partial x} = constant$ |
| Walls (Top, bottom, front and back) | | No-slip | $u = 0; \quad \frac{\partial p}{\partial n} = 0$ |
| Solid-fluid interface | Type 1 | No-slip | $u = 0; \quad \frac{\partial p}{\partial n} = 0$ |
| | Type 2 | No-slip | $u = 0; \quad \frac{\partial p}{\partial n} = 0$ |
| | Type 3 | Interior | No boundary condition |

*3.4 Tortuosity*

Tortuosity is a geometrical feature of porous media that quantify the straightness of the fluid channel formed by interconnecting pores. It is a dimensionless ratio of the length of the actual path to the length of the minimum path between two reference points in the flow-direction (refer Eq. A4 in appendix). In the absence of any obstacle, its value for a channel must be unity, however, in case of a porous medium, depending upon the morphology of the solid matrix, the value could be significantly larger than unity. In the present study, the tortuosity of the lattices is calculated by using the lengths of fluid trajectories in different regions and are mentioned in following table. Table 2 shows the tortuosity values of four lattices for void region. Moreover in Type 2 scenario, the solid region is treated as a fluid zone, and therefore, corresponding values of tortuosity are also calculated and shown in the same table.

Table 2: Tortuosity of two regions in the lattices.

| Sr. No. | Lattice name | Void region | Solid region (only for Type 2 scenario) |
|---|---|---|---|
| 1. | Diamond | 1.325 | 1.175 |
| 2. | I-WP | 1.051 | 1.123 |
| 3. | Primitive | 1.033 | 1.038 |
| 4. | Gyroid | 1.256 | 1.275 |

## 4. Solution Methodology

*4.1 Mesh generation process*

In the MS-Lattice software, four lattices with porosity 32% for each are generated and surface information is stored in STL format. The lattice surface is generated for unit cell of size 1 mm and repeated till 4 mm length in three directions. The STL file is then imported into the ANSYS ICEM-CFD software to create volumes of void and solid regions of porous lattice. The topology is properly checked at each step to ensure the connectivity of different surfaces and to avoid the negative volume during mesh generation. Unstructured tetrahedral mesh is employed for grid generation to deal with the complexity of the geometry as in figure below.

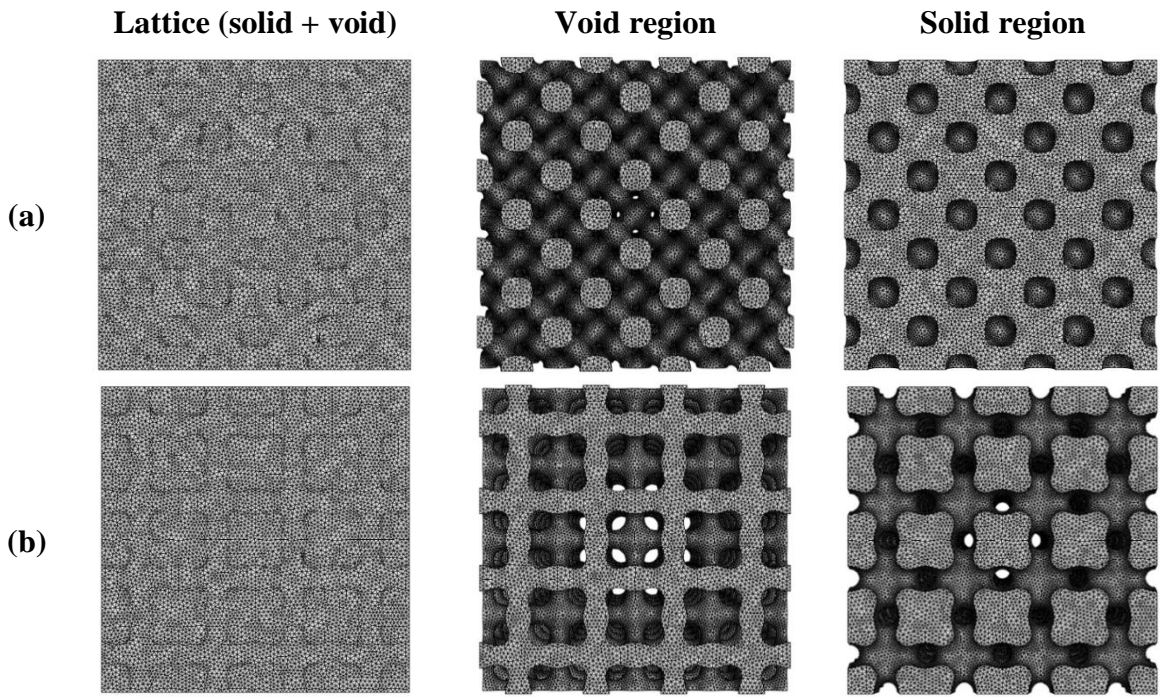

| | Lattice (solid + void) | Void region | Solid region |
|---|---|---|---|
| (a) | | | |
| (b) | | | |

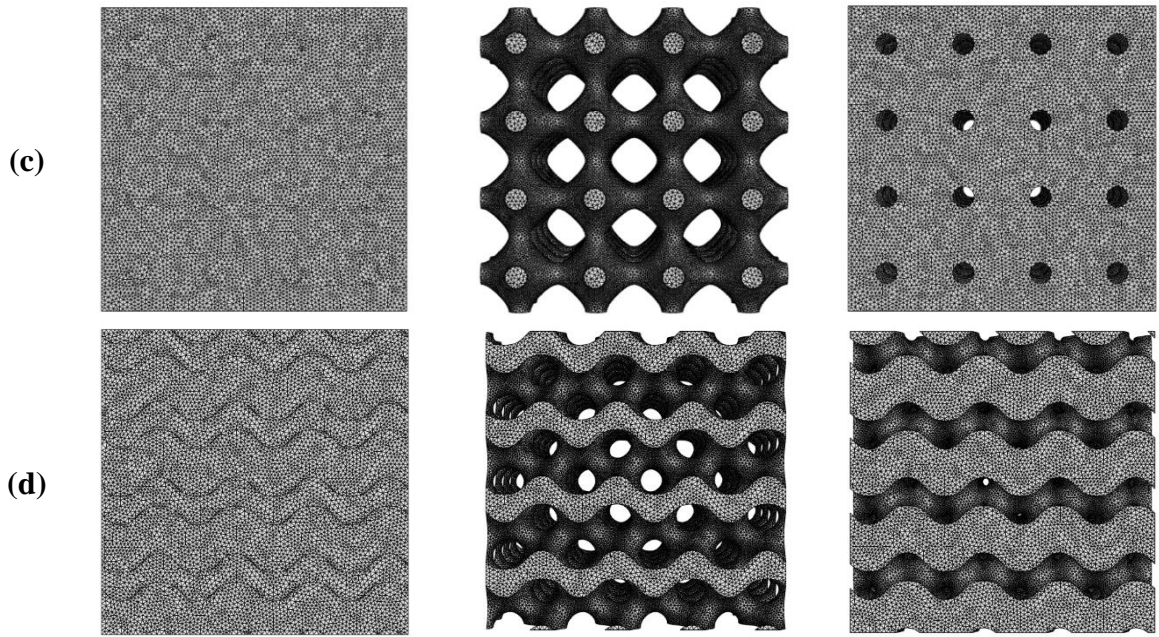

**Figure 5: Display of mesh in the void and solid regions for the four lattices, namely: (a) Diamond (b) I-WP (c) Primitive and (d) Gyroid.**

*4.2 Setting of simulation model*

In this work, hydrodynamics is studied for the flow of single-phase isothermal fluid in the porous lattices described in section 2. The 3-D steady state and incompressible governing equations are solved by using ANSYS® FLUENT with pressure-based solution approach. SIMPLE algorithm is used for pressure-velocity coupling, with standard scheme for discretization of the pressure and second order upwind scheme for the momentum. The iterations are performed up to the solution reached a convergence level of $10^{-6}$ for continuity and $10^{-8}$ for the x, y, and z-momentum equations respectively[45].

*4.3 Grid sensitivity tests*

The grid sensitivity is performed by taking four different meshes to verify the sensitivity of the grid density on the hydrodynamics for an empty channel. A non-uniform mesh where finer grids are used near the wall compared to the core to resolve the gradients with reasonable accuracy. The details of mesh used in this study are shown in Table 3.

**Table 3: Details of grids and results of grid dependency study.**

| Sr. No. | No. of nodes | Min. orthogonality | Max. aspect ratio | Pressure gradient (Pa/m) | % Difference |
|---|---|---|---|---|---|
| **Mesh 1** | 681 | 0.4 | 8.2 | -559.5 | - |
| **Mesh 2** | 3301 | 0.4 | 8.8 | -534.1 | 4.47 |
| **Mesh 3** | 47540 | 0.4 | 8.9 | -520.6 | 2.62 |
| **Mesh 4** | 63200 | 0.4 | 9.7 | -521.7 | 0.19 |

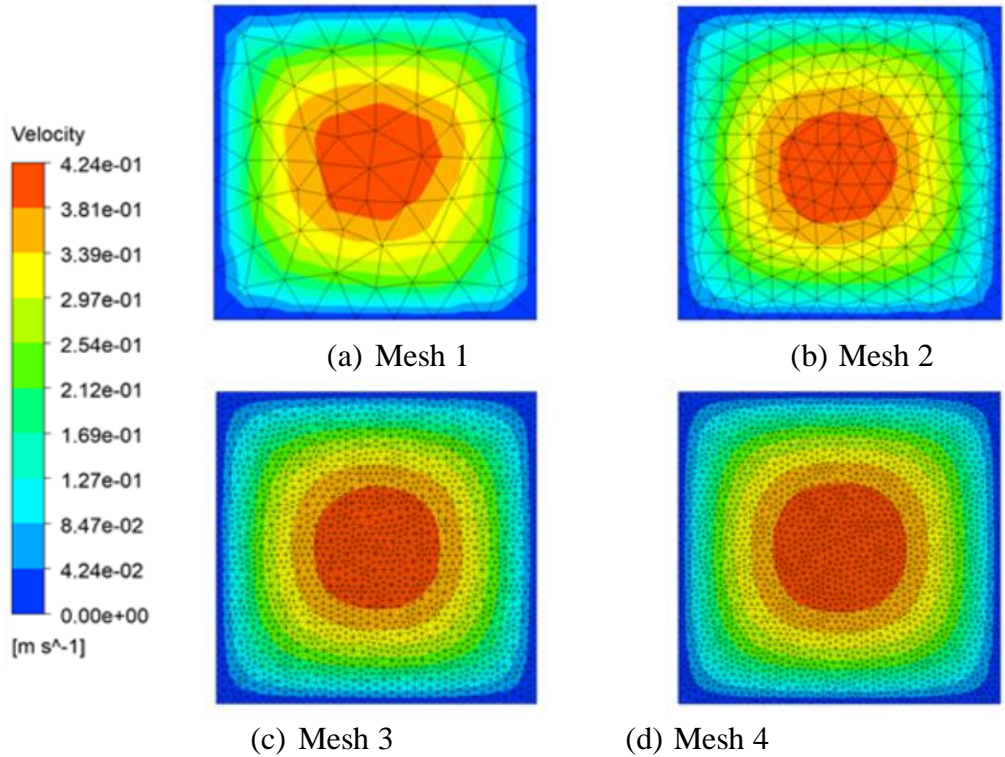

(a) Mesh 1      (b) Mesh 2

(c) Mesh 3      (d) Mesh 4

**Figure 6: Display of mesh and velocity contours for four different grids used in the grid sensitivity test at mass flow rate 4e-04 kg/s.**

The pressure gradients are compared for the mass flow rate at $4e-04\ kg/s$. In addition to that, velocity and axial pressure profiles are also compared as shown in Fig. 7. The difference in the value of pressure gradient between Mesh 1 and Mesh 2 is 4.47%, while between Mesh 2 and Mesh 3 is 2.62%, and Mesh 3 and Mesh 4 is 0.19%. The relative difference in the profiles of velocity and pressure variation is also negligibly small between the 'Mesh 3' grid and the 'Mesh 4'. Therefore, based on this test, Mesh 3 is chosen for further simulations in the present study.

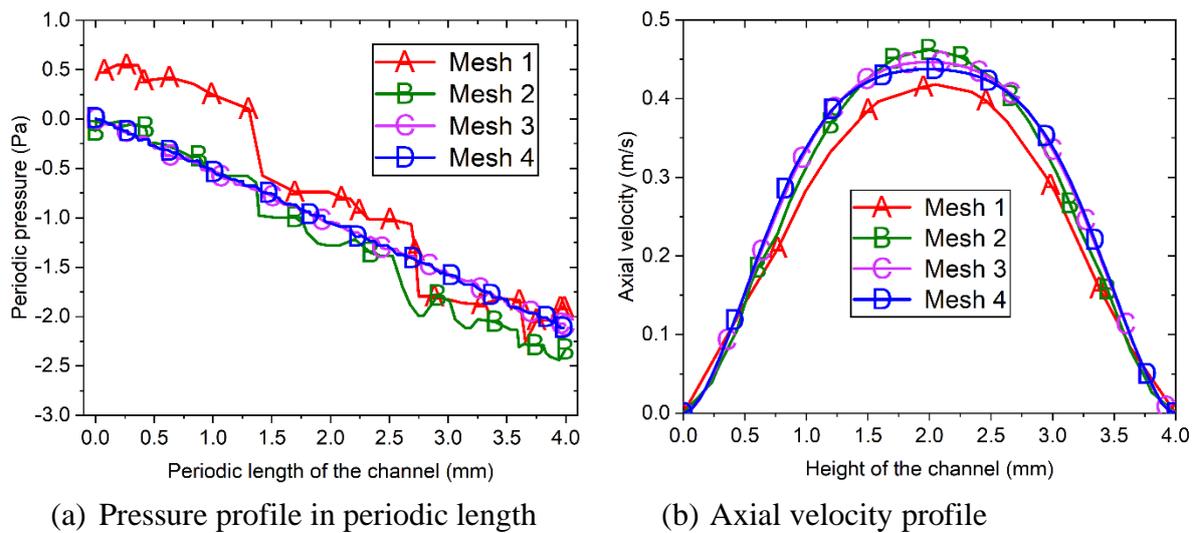

(a) Pressure profile in periodic length      (b) Axial velocity profile

**Figure 7: (a)Periodic pressure drops, and (b) velocity profiles for four different grids.**

*4.4 Validation of CFD model*

The above governing equation, boundary condition, and solution algorithm are validated for the flow in a mini square cross-section channel with periodic boundary condition between the inlet-outlet pair. The results of the numerical calculations are compared to empirical correlations given by Bahrami [46] and Shah [47] as shown in Eq. (9) and (10), respectively. For this pressure drop per unit length is plotted and compared for different Reynolds number, ranging from $Re = 0.01$ to $100$ (corresponding to $\dot{m} = 4e-08$ to $4e-04 \ kg/s$). The present results as shown in Fig. 8 match perfectly and therefore indicate the accuracy of the present numerical model.

$$\frac{\Delta p}{L} = \left(\frac{1}{3} - \frac{64}{\pi^5} \tanh\left(\frac{\pi}{2}\right)\right)^{-1} \frac{\mu}{\rho A^2} \dot{m} \tag{9}$$

$$\frac{\Delta p}{L} = \left(\frac{56.92}{2}\right) \frac{\mu}{\rho A^2} \dot{m} \tag{10}$$

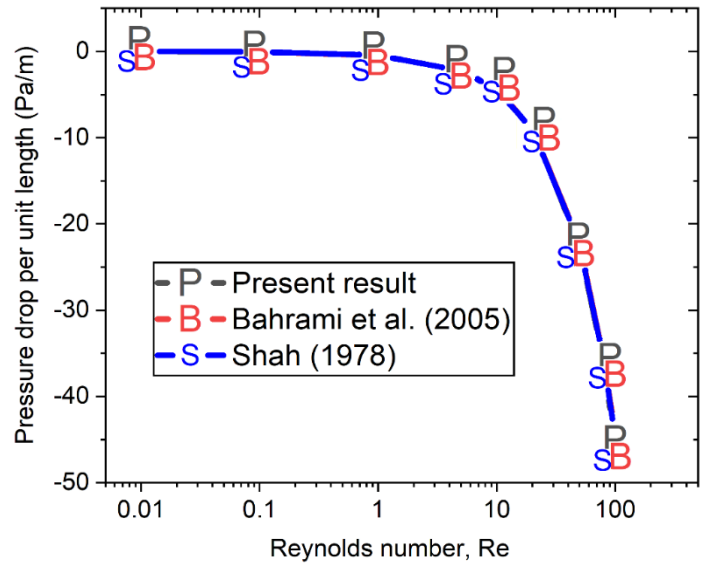

**Figure 8: Variation of pressure drop per unit length with Reynolds number.**

## 5. Results and Discussions

The pore-scale simulations have been performed for the flow in the TPMS lattice of twelve cases as described in the section 2 for mass flow rate ranging from $4e-08$ to $4e-04 \ kg/s$. The direct numerical simulations have been used for the first two types of scenarios, while, both direct numerical modelling for pore scale and macroscale modelling for the microporosity have been used for type 3 scenario. The detailed results of flow path, pressure drop, and velocity have been presented below.

Two very essential flow properties of a porous media, permeability and inertial drag factor are calculated and compared. For all lattices (Diamond, I-WP, Primitive, and Gyroid) and types (Type 1, 2, and 3) the pore-scale simulation is performed for different Reynolds number. The pressure drop is calculated for different mass flow rates and then compared with a general Darcy-Forchheimer model as shown in Eq. (11), to obtain the values of permeability and inertial drag factor. This pressure drop-mass flow rate relation, is used to characterize the flow through different types of porous media.

$$\frac{\Delta P}{L} = -\left(\frac{\mu}{\rho A K}\right)\dot{m} - \left(\frac{C_F}{\rho A^2 \sqrt{K}}\right)\dot{m}^2; \quad for\ \dot{m} = \rho A U \tag{11}$$

The permeability is calculated using the linear relationship between the pressure drop and mass flow rate (Darcian regime). Referring to Eq. (A2), it can be seen that the permeability is inversely proportional to the pressure gradient, i.e., a higher value of the slope of curve drawn between pressure gradient and mass flow rate gives the lesser value of the permeability. The curvature of the curve measures the inertial drag factor. The deviation of the curve from the Darcy line indicates contribution of inertial drag. Since, the overall nature of the curve is quadratic, a 2$^{nd}$ order curve fitting is performed in all cases and by the comparison of the coefficients with the standard Darcy-Forchheimer law, both permeability and inertial drag factor are obtained.

## 5.1 Type 1: Solid region is treated as solid matrix

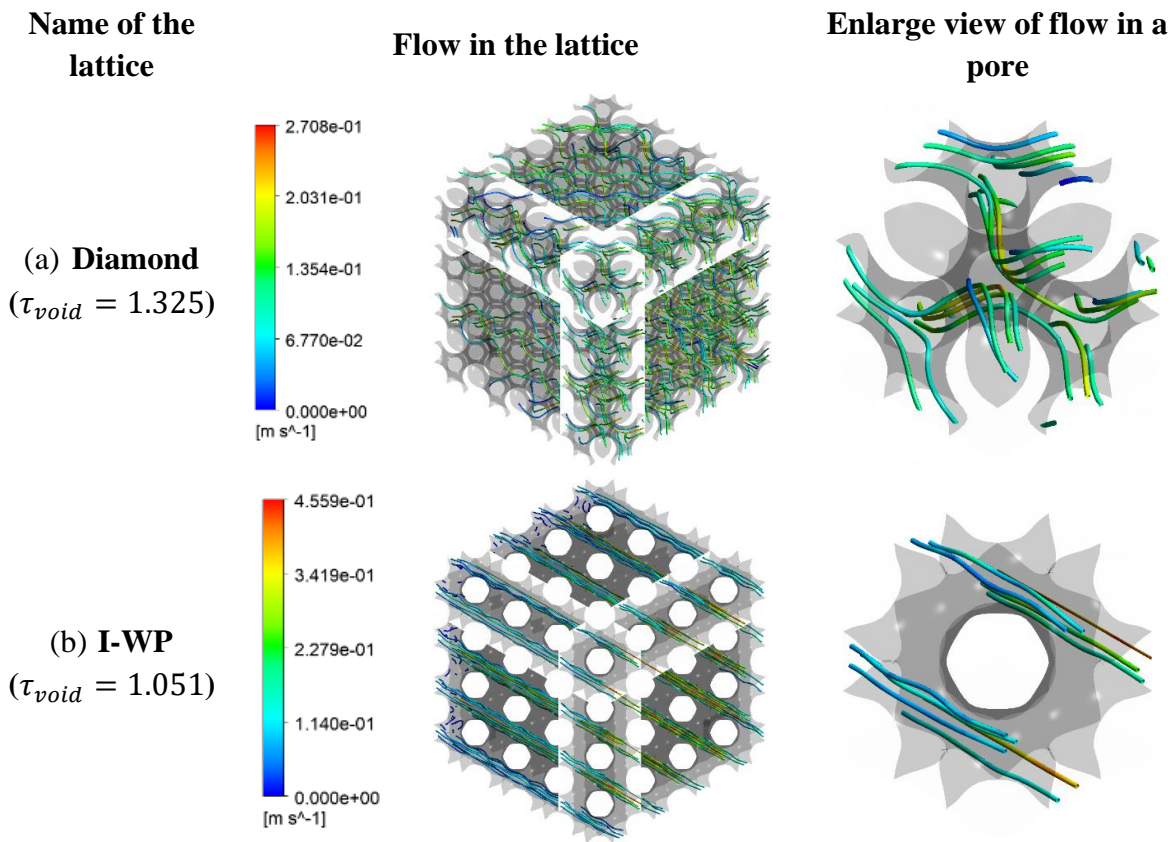

| Name of the lattice | Flow in the lattice | Enlarge view of flow in a pore |
|---|---|---|
| (a) **Diamond** ($\tau_{void} = 1.325$) | | |
| (b) **I-WP** ($\tau_{void} = 1.051$) | | |

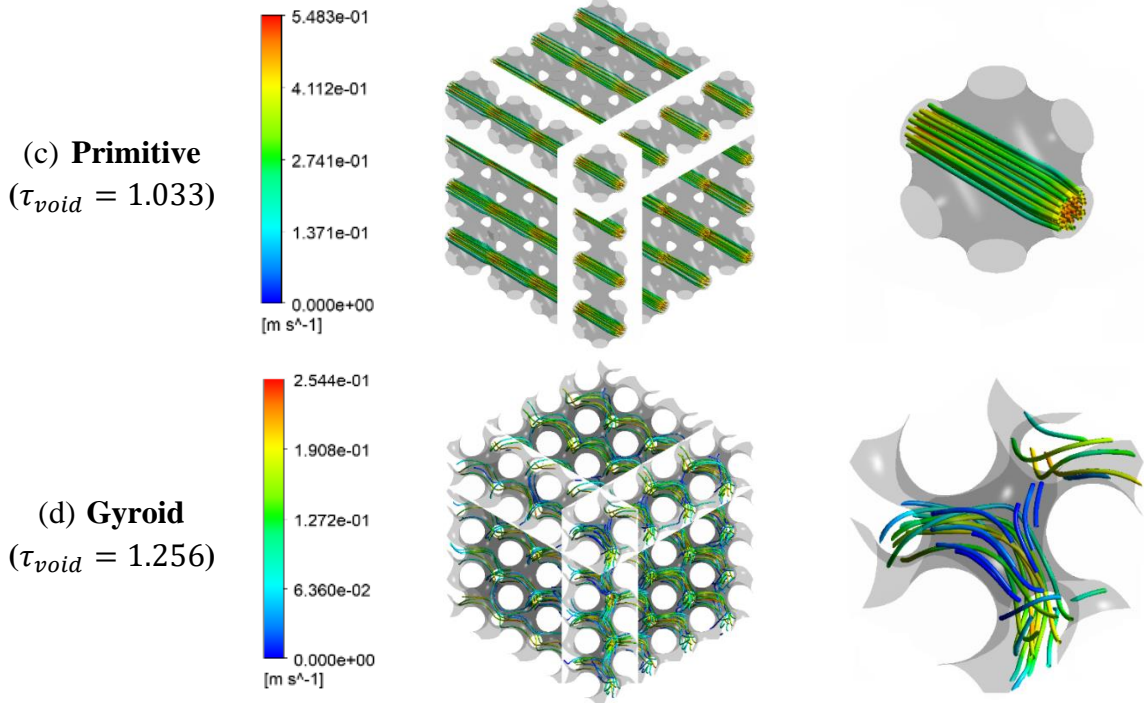

**Figure 9: Flow trajectories in four lattices and in a pore for Type 1 scenario for mass flow rate 4e-04 kg/s.**

In this case, the solid region of lattice is treated as solid matrix and the fluid flow is only through the void zone which is 32% (porosity ε = 0.32) of the total cubical volume. Fig. 9 shows the fluid path in the four lattices using flow trajectories at mass flow rate 4e-4 kg/sec. In the Diamond lattice (Fig. 9(a)), the fluid path is observed to be most tortuous with its tortuosity value being maximum as 1.325. In case of I-WP lattice (Fig. 9(b)), the fluid path is relatively less tortuous with value of tortuosity as $\tau = 1.051$. In Primitive lattice (Fig. 9(c)), again the fluid path is looking to be quite straighter and its tortuosity is calculated to be minimum as $\tau = 1.033$. Lastly, in Gyroid lattice (Fig. 9(d)), the flow path is observed to be very distorted and takes their shape as of helix with tortuosity value as $\tau = 1.256$. The lattices with high tortuosity i.e., Diamond and Gyroid has relatively low value of pore velocity 0.27 and 0.25 m/s respectively, whereas the lattices with low tortuosity i.e., I-WP and Primitive has relatively high pore velocity i.e., 0.45 and 0.54 m/sec respectively, as shown in Fig. 9 and also in Table 5.

The drop in pressure with different flow rates is shown in Fig. 10 along with the percentage deviation from the linear (or Darcy) model. For the Diamond lattice, the linear relationship is retained only up to mass flow rate, $\dot{m} = 4e - 05\ kg/s$ (corresponding Re = 10) and after that inertial (non-linear) term becomes significant. At mass flow rate of $4e - 04\ kg/s$ (Re = 100), the pressure gradient is -74.19 kPa/m and maximum deviation in this case is under 20%. In I-WP lattice, the linear behavior is followed till mass flow rate, $\dot{m} = 3.2e - 05\ kg/s$ (corresponding Re = 8) and the pressure gradient at mass flow rate, $\dot{m} = 4e - 04\ kg/s$ is -100.28 kPa/m with maximum value of deviation under 30%. In the Primitive lattice, linear behavior is followed for comparatively small value of mass flow rate, $\dot{m} = 1.6e - 05\ kg/s$ (corresponding Reynolds number Re = 4), and the deviation from the Darcy line can be observed to grow swiftly from there as seen by higher percentage deviation (> 40%). The pressure gradient in this case, -76 kPa/m is recorded for mass flow rate of $\dot{m} = 4e - 4\ kg/s$ (corresponding Re = 100). Lastly, in Gyroid lattice, the linear behavior is observed till mass flow rate of $\dot{m} = 4e - 05\ kg/s$ and corresponding Reynolds number Re = 10, and the deviation is observed to be growing quickly and showing a very large deviation (~60%). The pressure gradient at mass flow rate 4e-4 kg/sec (corresponding Re = 100) is computed to be -58.35 kPa/m.

Table 6 depicts the permeability and inertial drag factor for all lattices of Type 1 treatment for the solid region. The I-WP lattice shows the minimum Darcy number as $Da = 1.9e - 05$, followed by Diamond and Primitive lattices as $2.7e - 05$ and $3.2e - 05$ respectively, and maximum for the Gyroid lattice as $Da = 4.9e - 05$. It is worth to notice that the lower value of permeability indicates a higher viscous resistance, whereas higher value of inertial drag factor indicates a higher form (or inertial) resistance. Therefore, I-WP lattice offers higher viscous and lower inertial resistance, however, opposite of this is found for Gyroid lattice. Table 7 shows the maximum velocity and pressure gradient for different mass flow rates through these lattices. The linear behavior between pressure gradient and flow rate (indicated here using Reynolds number) is retained up to $\dot{m} = 4\text{e} - 05$ kg/s (or $Re = 10$), commonly for all the lattices.

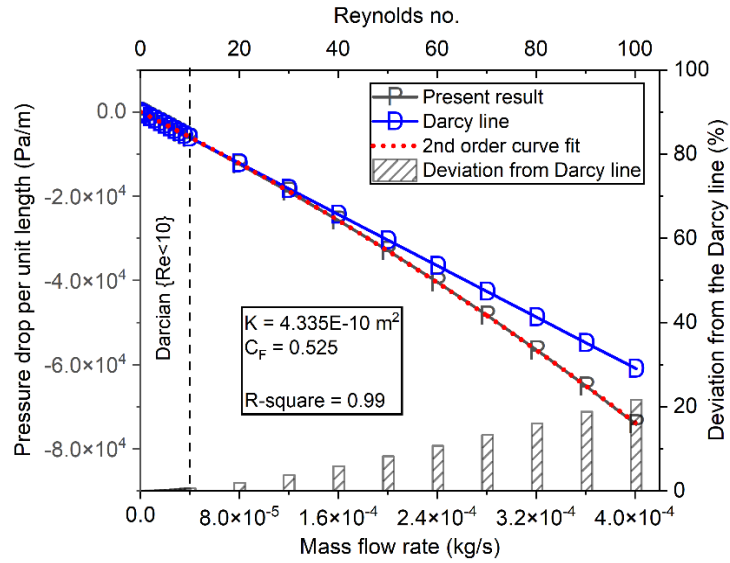 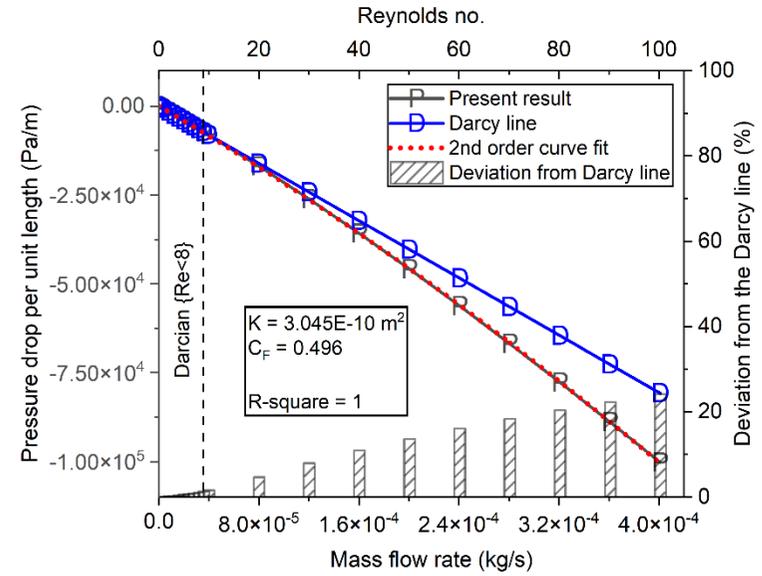
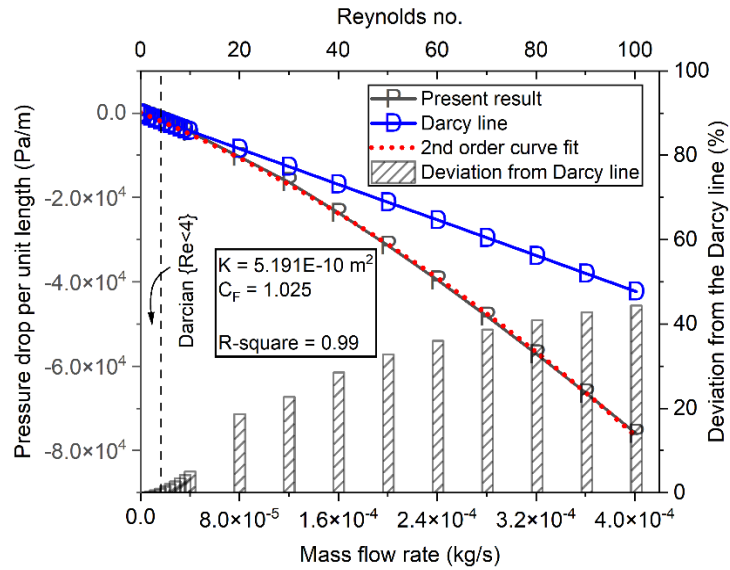 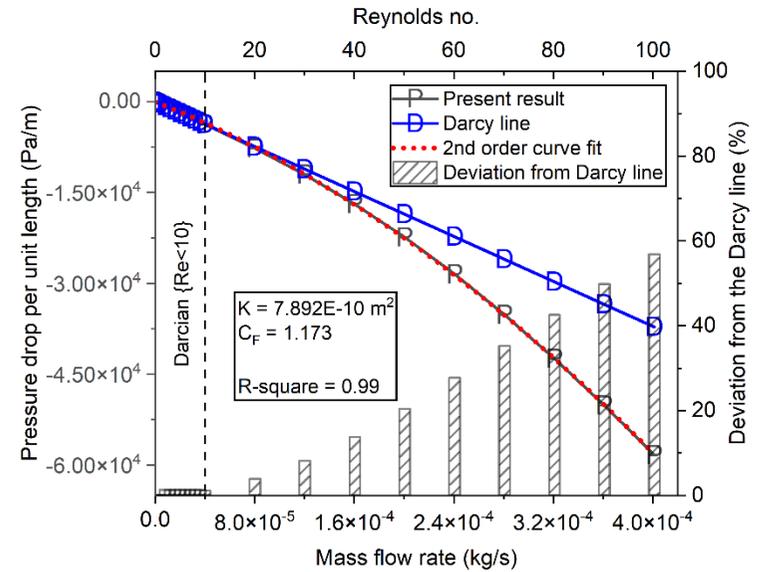

**Figure 10: Variation of pressure drop per unit length with mass flow rates for lattices with Type 1 scenario for the solid region.**

**Table 4: Porous media characteristics for Type 1 treatment of solid region.**

| Lattice name | Porosity, $\varepsilon$ | Permeability, K | Darcy number, Da | Inertial drag factor, $C_F$ |
|---|---|---|---|---|
| **Diamond** | 0.32 | $4.335 \times 10^{-10}$ m$^2$ | $2.7 \times 10^{-5}$ | 0.525 |
| **I-WP** | 0.32 | $3.045 \times 10^{-10}$ m$^2$ | $1.9 \times 10^{-5}$ | 0.496 |
| **Primitive** | 0.32 | $5.191 \times 10^{-10}$ m$^2$ | $3.2 \times 10^{-5}$ | 1.025 |
| **Gyroid** | 0.32 | $7.892 \times 10^{-10}$ m$^2$ | $4.9 \times 10^{-5}$ | 1.173 |

**Table 5: Maximum velocity and pressure gradient for different mass flow rates for Type 1 treatment of solid region.**

| | **Diamond** | | **I-WP** | | **Primitive** | | **Gyroid** | |
|---|---|---|---|---|---|---|---|---|
| $\dot{m}$ (kg/s) | Max. velocity (m/s) | Pressure gradient (kPa/m) | Max. velocity (m/s) | Pressure gradient (kPa/m) | Max. velocity (m/s) | Pressure gradient (kPa/m) | Max. velocity (m/s) | Pressure gradient (kPa/m) |
| **4e-8** | 2.8e-05 | -6.1e-03 | 5.3e-05 | -8.1e-03 | 6.7e-05 | -4.2e-03 | 2.6e-05 | -3.7e-03 |
| **4e-7** | 2.8e-04 | -6.1e-02 | 5.3e-04 | -8.1e-02 | 6.7e-04 | -4.2e-02 | 2.6e-04 | -3.7e-02 |
| **4e-6** | 2.8e-03 | -6.1e-01 | 5.3e-03 | -8.1e-01 | 6.7e-03 | -4.2e-01 | 2.6e-03 | -3.7e-01 |
| **4e-5** | 2.8e-02 | -6.1 | 5.2e-02 | -8.2 | 6.6e-02 | -4.4 | 2.6e-02 | -3.7 |
| **4e-4** | 2.7e-01 | -74.2 | 4.5e-01 | -100.3 | 5.4e-01 | -76 | 2.5e-01 | -58.3 |

The pressure drop monotonically increases with the mass flow rate for all the lattices. As shown in Fig. 11, the I-WP and Gyroid lattice have respectively highest and lowest pressure drop for all mass flow rate studied in the present work. This indicates the I-WP lattice offer maximum resistance to flow followed by Diamond and Primitive lattices, and minimum for the Gyroid lattice.

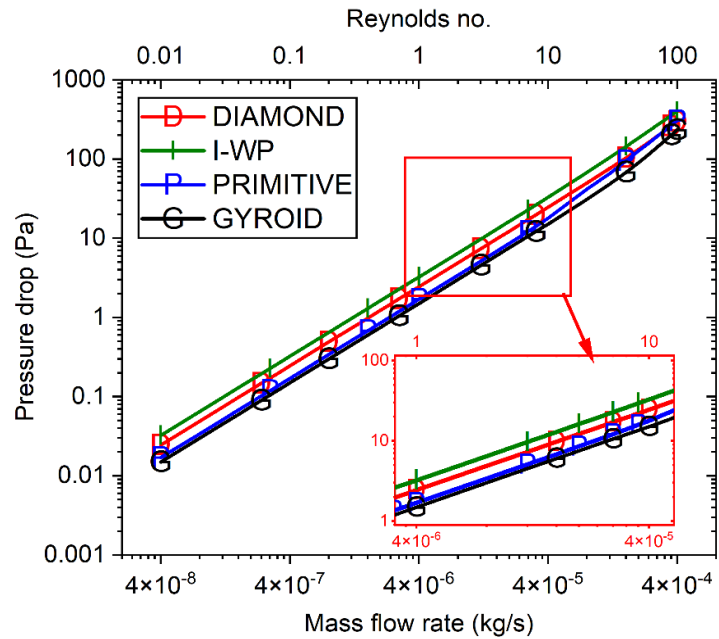

**Figure 11: Variation of pressure drop with mass flow rate for lattices with Type 1 scenario.**

*5.2 Type 2: Solid region is treated as fluid zone*

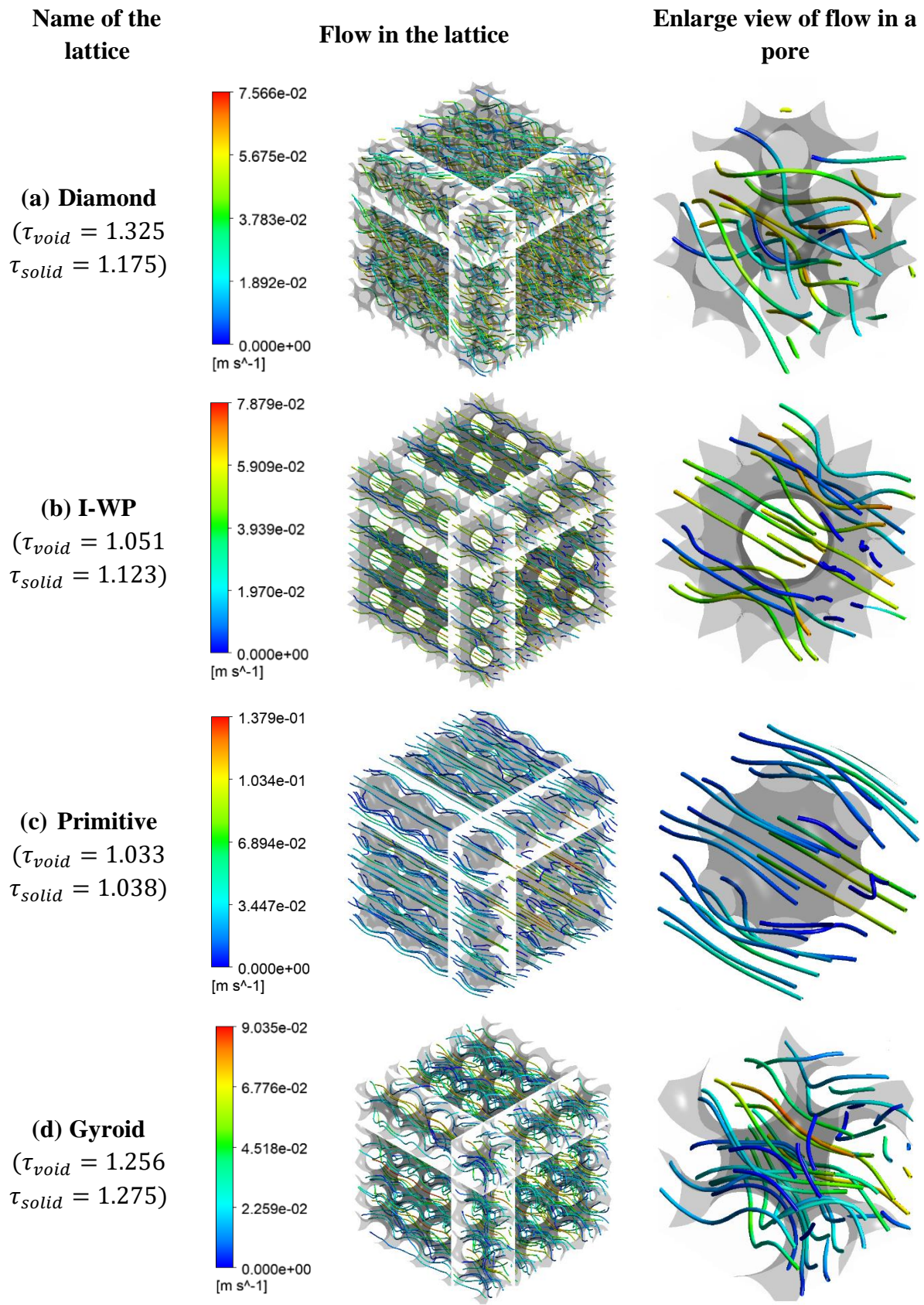

**Figure 12: Flow trajectories in four lattices and their unit cell or pore for Type 2 scenario for mass flow rate 4e-04 kg/s.**

For this type, solid region of the original lattices is converted in the fluid zone, thus, the two flow regions are separated by a zero-thickness wall which may break the boundary layer growth. Therefore, almost the entire volume is effectively open for the flow, and this type of super-porosity is common in foams, particularly metal foams. Fig. 12 depicts the fluid trajectories for the lattices with Type 2 treatment of the solid region. In case of Diamond (Fig. 12(a)), due to the distorted shape of the interstitial wall which separates the two flow regions ($\tau_{void} = 1.325$ and $\tau_{solid} = 1.175$), the overall flow path is also tortuous in shape. In I-WP lattice (Fig. 12(b)), the flow path is observed to be following a combination of straight ($\tau_{void} = 1.051$) and helical path ($\tau_{solid} = 1.123$) for two regions. In Primitive lattice (Fig. 12(c)), the flow path for void region is found to be relatively straighter in both regions ($\tau_{void} = 1.033$ and $\tau_{solid} = 1.038$). Lastly in Gyroid lattice (Fig. 12(d)), for both regions; solid and void, the fluid paths are found to be following helical trajectories ($\tau_{void} = 1.256$ and $\tau_{solid} = 1.275$). The straightness of fluid trajectory also indicates the lower reduction in the pore velocity of maximum value, therefore it is observed in Fig. 12 and also in Table 7 that the lattices with overall high tortuosity i.e., Diamond, I-WP and Gyroid has relatively low value of pore velocity of maximum value, whereas, the Primitive lattice has relatively high pore velocity of maximum value.

The variation of pressure gradient with the mass flow rate along with the deviation in the curve measured with reference to the Darcy line is shown in Fig. 13. In case of Diamond lattice, the linear behavior of pressure drop is observed up to mass flow rate, $\dot{m} = 8e - 05\ kg/s$ and corresponding Reynolds number $Re = 20$ and the pressure gradient for $\dot{m} = 4e - 04\ kg/s$ flow rate is observed as -7.7 kPa/m, which is 10 times lower than the Type 1 Diamond lattice. Overall, deviation of curve from the Darcy line is not significant even for higher mass flow rates. The linear behavior in pressure drop is followed up to $\dot{m} = 4e - 05\ kg/s$ or (corresponding $Re = 10$) after which it deviates with a slow pace, in I-WP lattice. Moreover, the pressure gradient for $\dot{m} = 4e - 04\ kg/s$ is -6 kPa/m, which is almost 16 times lower than the Type 1 I-WP lattice. Furthermore, in Primitive lattice, the linear behavior is followed up to $\dot{m} = 4e - 05\ kg/s$ or corresponding $Re = 10$, and after which, the deviation progress slowly with increasing flow rate. Pressure gradient at $\dot{m} = 4e - 04\ kg/s$ is -3.3 kPa/m, which is almost 20 times lower than its Type 1 counterpart. Lastly for Gyroid lattice, linear behavior is followed till mass flow rate $\dot{m} = 4e - 05\ kg/s$ ($or\ Re = 10$), and again the deviation from linearity is not significantly large. The pressure gradient at $\dot{m} = 4e - 04\ kg/s$ is calculated to be -5.6 kPa/m, almost 10 times lower than compared to its Type 1 counterpart. Overall observation suggests that in Type 2 case, the inertial drag factors are comparatively low and permeability are one-order higher than Type-1 case which leads to lower drop in pressure.

Table 6 shows Darcy number and inertial drag factor for Type 2 treatment of lattices. Darcy number has minimum value ($2.4e - 04$) for Diamond lattice, followed by I-WP and Gyroid (both $3.6e - 04$), and has the maximum value ($4.7e - 04$) for Primitive lattice. However, the drag factors are not in same trend. The value of inertial drag factor is minimum as 0.074 for Primitive lattice, which is followed by Diamond and Gyroid with values 0.119 and 0.155 respectively and has maximum value calculated for I-WP lattice as 0.202.

Table 7 presents the maximum velocity and pressure gradient for Type 2 lattices. Once again it is observed that linear behavior of the pressure drop against flow rate is retained only up to $\dot{m} = 4e - 05\ kg/s\ (or\ Re = 10)$, and the Darcy law can conveniently be applied for the lattices to model their flow behavior in Type 2 settings. On comparing the pore velocity for mass flow rate of $4e - 04\ kg/s$, it is observed that that the largest value of velocity is obtained for Primitive lattice, which may be due to its comparatively smaller tortuosity by which it offers lower resistance (since it also has maximum permeability and minimum drag factor) to the flow. Furthermore, Diamond and I-WP lattices are showing nearly similar value of max. velocity to be 0.075 and 0.078 m/s respectively. Lastly, the Gyroid lattice is showing smallest value of pore velocity (0.09 m/s) at the same mass flow rate i.e., $\dot{m} = 4e - 04\ kg/s$. In comparison with the Type 1 lattice, here in Type 2 lattice pressure drops are significantly reduced and the Darcy number is increased by one-order.

**Table 6: Details of porous media properties for Type 2 treatment of solid region.**

| Lattice name | Porosity, ε | Permeability, K | Darcy number, Da | Inertial drag factor, $C_F$ |
|---|---|---|---|---|
| **Diamond** | ~1 | $3.898 \times 10^{-9}\ m^2$ | $2.4 \times 10^{-4}$ | 0.119 |
| **I-WP** | ~1 | $5.772 \times 10^{-9}\ m^2$ | $3.6 \times 10^{-4}$ | 0.202 |
| **Primitive** | ~1 | $7.505 \times 10^{-9}\ m^2$ | $4.7 \times 10^{-4}$ | 0.074 |
| **Gyroid** | ~1 | $5.761 \times 10^{-9}\ m^2$ | $3.6 \times 10^{-4}$ | 0.155 |

**Table 7: Maximum velocity and pressure gradient for different Reynolds number for Type 2 treatment of solid region.**

| | Diamond | | I-WP | | Primitive | | Gyroid | |
|---|---|---|---|---|---|---|---|---|
| $\dot{m}$ (kg/s) | Max. velocity (m/s) | Pressure gradient (kPa/m) | Max. velocity (m/s) | Pressure gradient (kPa/m) | Max. velocity (m/s) | Pressure gradient (kPa/m) | Max. velocity (m/s) | Pressure gradient (kPa/m) |
| **4e-8** | 7.1e-06 | -6.7e-04 | 7.1e-06 | -4.6e-04 | 1.5e-05 | -3.3e-04 | 8.7e-06 | -4.4e-04 |
| **4e-7** | 7.1e-05 | -6.7e-03 | 7.1e-05 | -4.6e-03 | 1.5e-04 | -3.3e-03 | 8.7e-05 | -4.4e-03 |
| **4e-6** | 7.1e-04 | -6.7e-02 | 7.1e-04 | -4.6e-02 | 1.5e-03 | -3.3e-02 | 8.7e-04 | -4.4e-02 |
| **4e-5** | 7.1e-03 | -6.7e-01 | 7.1e-03 | -4.6e-01 | 1.5e-02 | -3.3e-01 | 8.7e-03 | -4.5e-01 |
| **4e-4** | 7.5e-02 | -7.7 | 7.8e-02 | -6.1 | 1.4e-01 | -3.8 | 9.0e-02 | -5.6 |

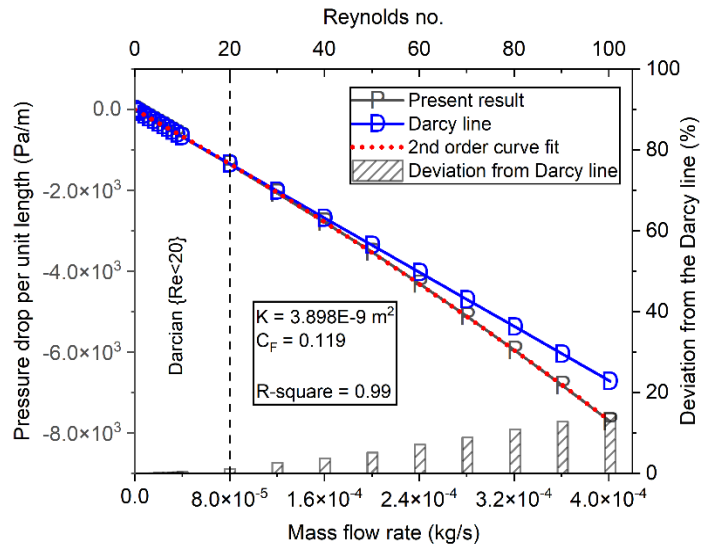
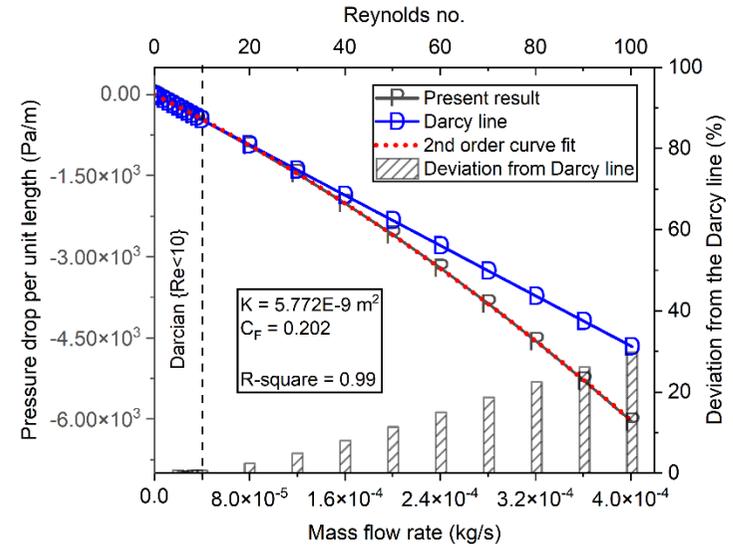
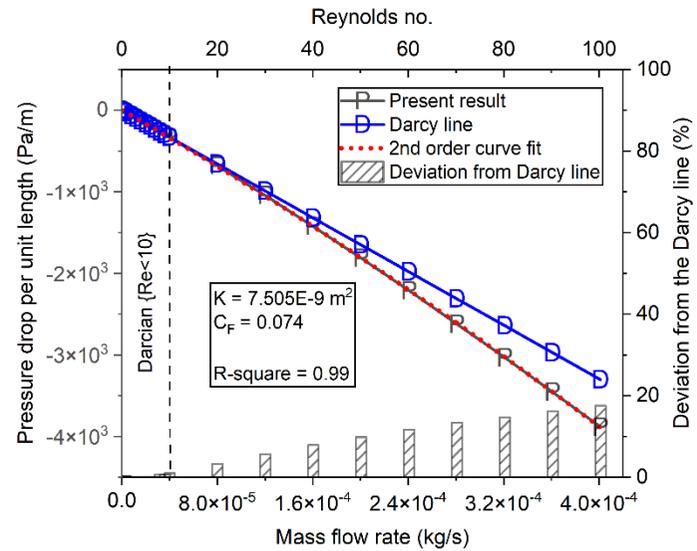
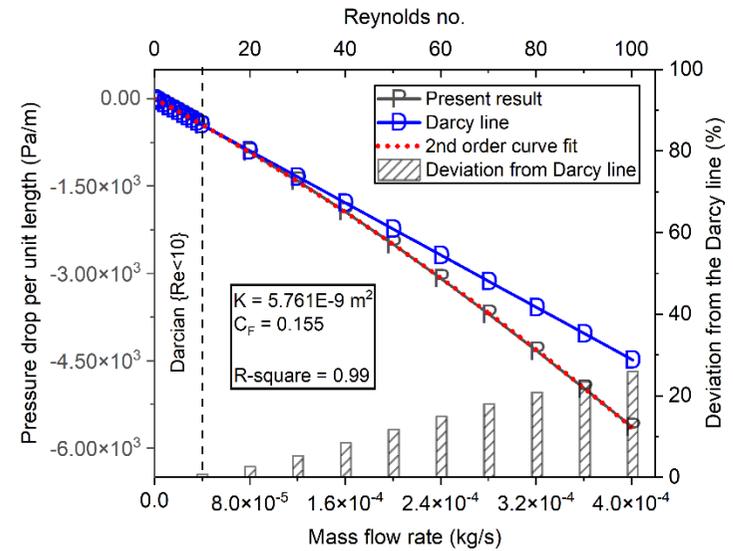

**Figure 13: Variation of pressure drop per unit length with mass flow rates for lattices with Type 2 scenario for the solid region.**

The variation of pressure drop with Reynolds number for all the lattices with Type 2 treatment of solid region is depicted in Fig. 14. The Diamond and Primitive lattices show the highest and the lowest pressure drops respectively for the entire range of the Reynolds number, which means that their resistance to pass the fluid through the tortuous path provided by the complicated thin-walled interior are respectively maximum and minimum. This is also clearly visible in their permeability values. For the remaining two lattices; I-WP and Gyroid, the drop in pressure or flow resistance is almost same for the entire range of flow rates.

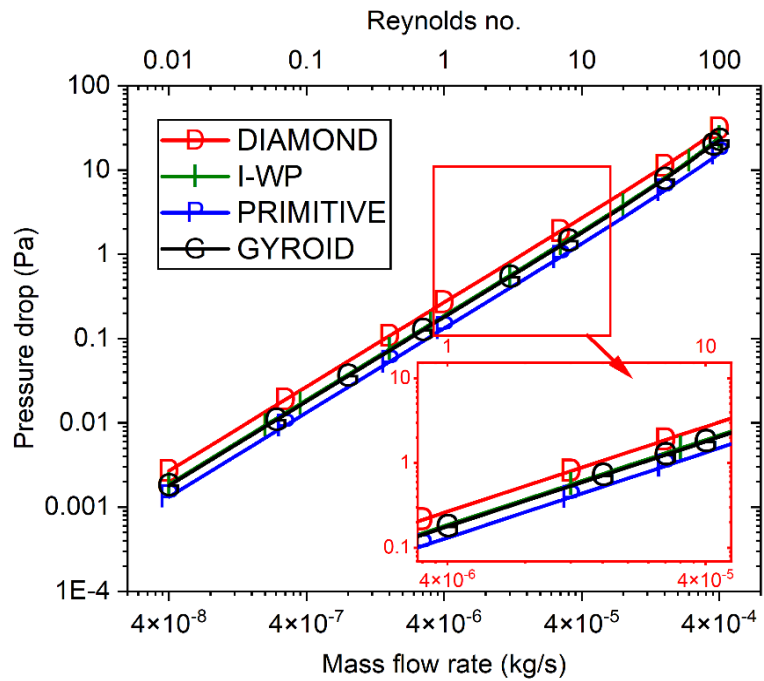

Figure 14: Variation of pressure drop with Reynolds number for lattices with Type 2 scenario.

*5.3 Type 3: Solid region is treated as porous zone*

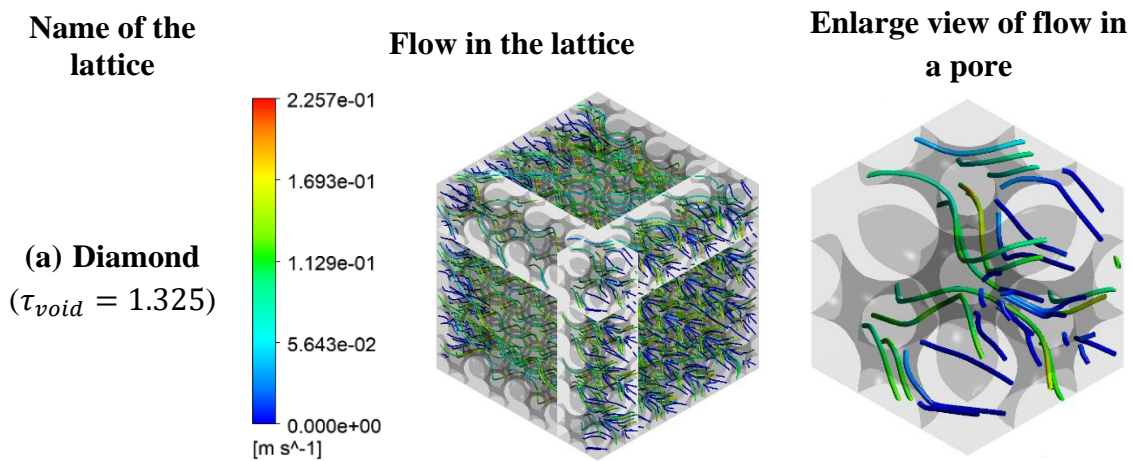

| Name of the lattice | Flow in the lattice | Enlarge view of flow in a pore |
|---|---|---|
| (a) Diamond ($\tau_{void} = 1.325$) | | |

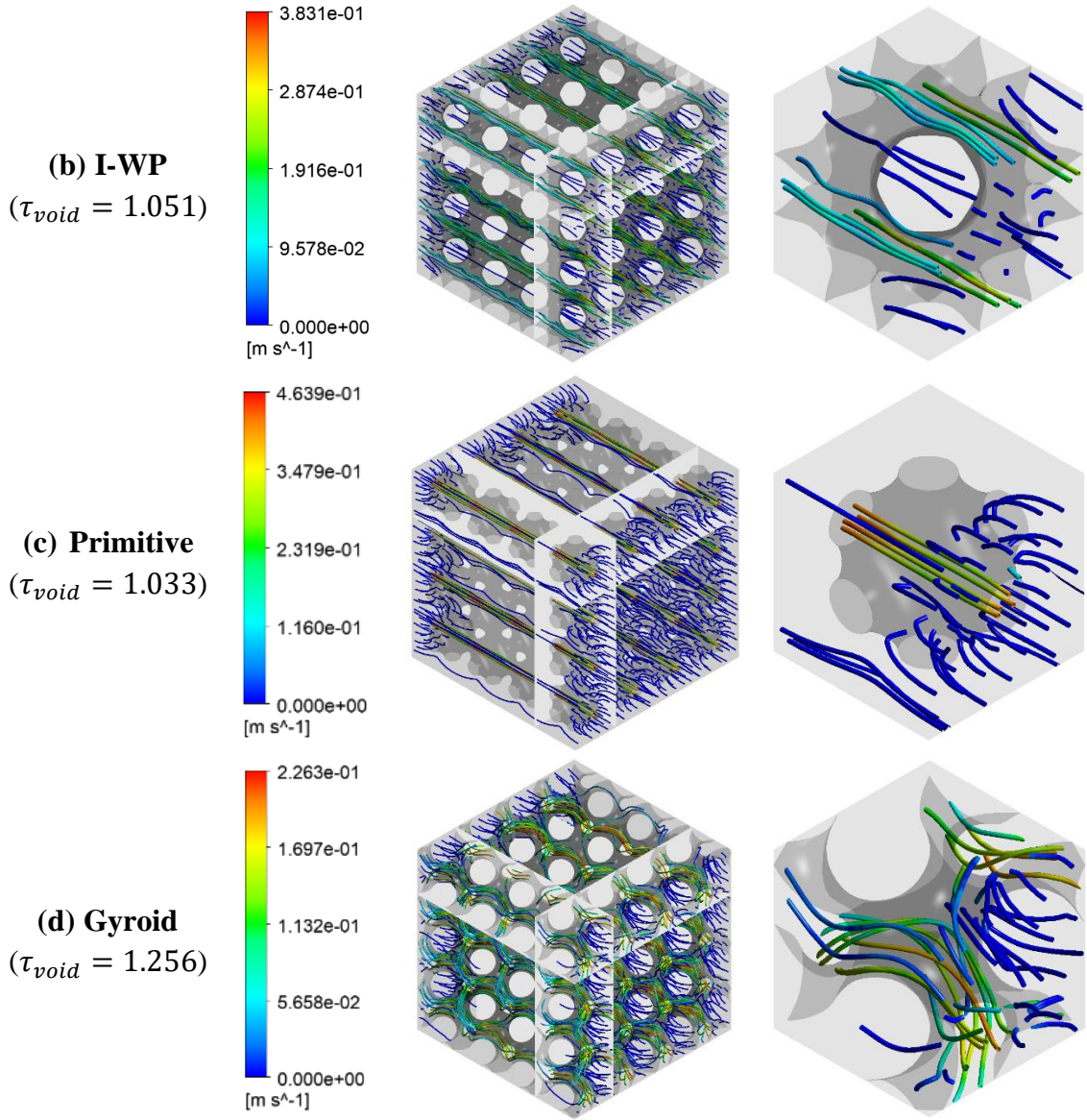

**Figure 15: Flow trajectories in four lattices and their unit cell or pore for Type 3 scenario for mass flow rate 4e-04 kg/s.**

In this case, the solid region of the lattices is considered as the porous zone with porosity 32% with the consideration that the pore size in the solid zone is very small (microporosity) compared to the pore scale of the lattice. The direct numerical simulation is performed for the pore scale of the lattice while porous media modelling (Darcy-Forchheimer-Brinkman model) is considered for the microporosity. Furthermore, the surfaces between the pore of lattice (void zone) and porous medium (microporous zone) is made permeable by treating it as interior and allow the fluid to pass through it. The simulation results for this scenario are presented below.

The trajectories of fluid in the lattices at mass flow rate of $4e - 04 \, kg/s$ (or corresponding $Re = 100$) is illustrated in Fig. 15. Two different regions are present in the computational domain, first region is voids or cavity in the lattice in which the fluid flows without additional resistance, whereas, the second is solid region assigned as the porous zone with induced viscous and inertial resistances caused due to micro-porosity. The trajectory of the fluid path in the void region is helical in Diamond lattice, however in its microporous region, the flow is damped down and thus magnitude of velocity is comparatively small. Further, the shape of flow path in the void region is straight line, however, the flow path is zig-zag in the microporous zone of I-WP lattice. Furthermore, in Primitive lattice, the shape of flow path is similar to the previous (I-WP lattice), in which the flow paths are straight and zig-zag, respectively, in the both the void region and microporous (porous zone) making it to have less tortuous. Lastly, flow path in the void region takes a helical shape in Gyroid lattice, and porous zone shows a damped effect similar to the Diamond lattice and therefore overall nature of the flow path is tortuous. The tortuosity in case of Type 3 lattice is similar to the Type 1 as far as the void region is considered, however, it is not calculatable for the microporous region due to porous media flow modelling is used. Similar to the Type 1, again the lattices with high tortuosity i.e., Diamond and Gyroid has relatively low value of pore velocity $2.2e - 01\ m/s$ in both, whereas, the lattices with low tortuosity i.e., I-WP and Primitive has relatively high pore velocity i.e., $3.8e - 01$ and $4.6e - 01\ m/s$ respectively, as shown in Fig. 15 and also in Table 9.

The variation of pressure gradient with the mass flow rate, pressure gradient for Darcy law, and deviation between them in that flow range is presented in Fig. 16. The linear behavior of pressure gradient with flow rate is observed up to $\dot{m} = 8e - 05\ kg/s$ or corresponding $Re = 20$ and after which the deviation grows at a slower pace for Diamond lattice. The pressure gradient in this, at mass flow rate, $\dot{m} = 4e - 04\ kg/s$ (Re = 100) is equal to -54.9 kPa/m. Further, the linear behavior is followed till mass flow rate, $\dot{m} = 3.2e - 05\ kg/s$ or corresponding $Re = 8$ in I-WP lattice and after this a deviation is observed to be growing slowly with higher flow rate. The pressure gradient at $\dot{m} = 4e - 04\ kg/s$ (or Re = 100) flow rate is equal to -72.7 kPa/m. Further, linear behavior is followed only till $\dot{m} = 1.6e - 05\ kg/s$ or corresponding $Re = 4$ in Primitive lattice, but after this percentage deviation grows at a very fast rate making it to reach 75% for $\dot{m} = 4e - 04\ kg/s$ or corresponding $Re = 100$ case. The pressure gradient in $\dot{m} = 4e - 04\ kg/s$ case is -56.2 Pa/m. Finally, in Gyroid lattice, the linear behavior is retained only till $\dot{m} = 3.6e - 05\ kg/s$ or corresponding $Re = 9$, after which it deviates sharply. The pressure gradient at flow rate $\dot{m} = 4e - 04\ kg/s$ is equal to -45.4 kPa/m.

The Darcy number and inertial drag factor for all lattices are presented in Table 8 in Type 3 treatment of solid region. The Darcy number is minimum for the I-WP lattice ($Da = 2.6e - 05$), followed by Diamond and Primitive lattices ($3.1e - 05$ and $4.3e - 05$ respectively) and has the maximum value $Da = 6.1e - 05$ for the Gyroid lattice. The inertial drag factor is minimum of value 0.405 for Diamond lattice, followed by the I-WP and Primitive with values 0.407 and 0.851 respectively and is maximum for the Gyroid lattice with value 0.961.

The maximum velocity and axial pressure gradient for different orders of mass flow rate are shown in the Table 9. Due to the additional resistances offered by the micro-porosity present in the solid region and which is creating dampening effect to the inertial forces, the linear relation between pressure drop and mass flow rate is strongly followed till mass flow rate, $\dot{m} = 4e - 06\ kg/s$ or corresponding $Re = 1$ and continues till $\dot{m} = 4e - 05\ kg/s$ or $Re = 10$ with insignificant errors.

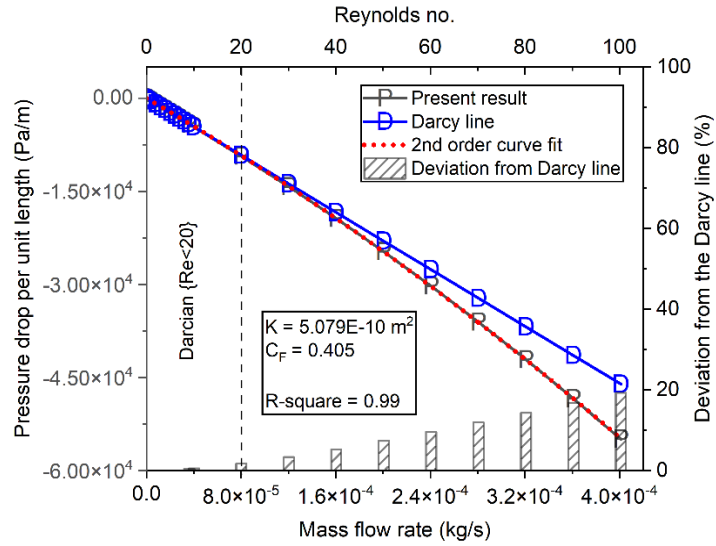 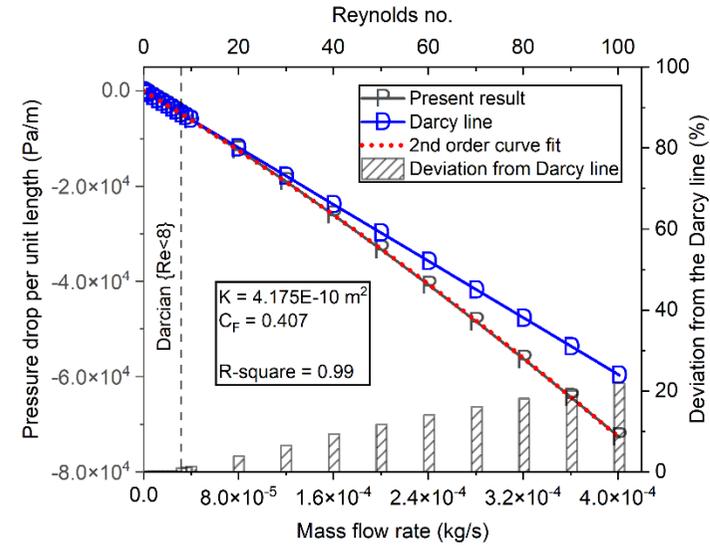
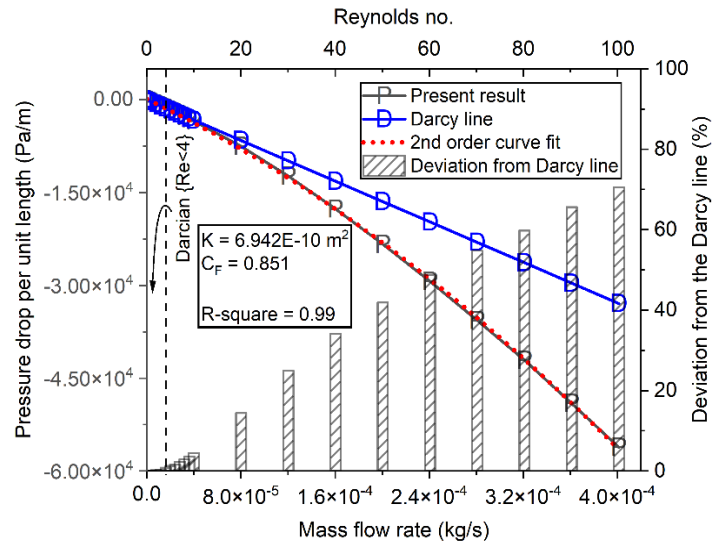 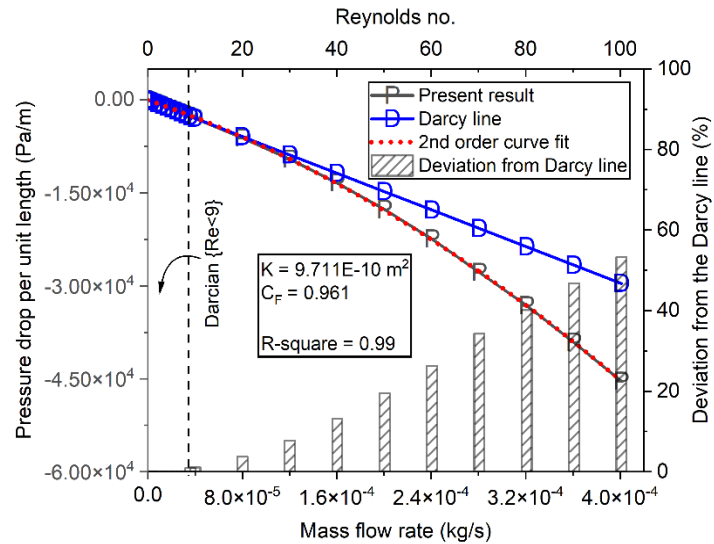

**Figure 16: Variation of pressure drop per unit length with mass flow rates for lattices with Type 3 scenario for the solid region.**

**Table 8: Details of porous media properties for Type 3 treatment of solid region.**

| Lattice name | Porosity, ε | Permeability, K | Darcy number, Da | Inertial drag factor, $C_F$ |
|---|---|---|---|---|
| **Diamond** | 0.54 | $5.079 \times 10^{-10}$ m² | $3.1 \times 10^{-5}$ | 0.405 |
| **I-WP** | 0.54 | $4.175 \times 10^{-10}$ m² | $2.6 \times 10^{-5}$ | 0.407 |
| **Primitive** | 0.54 | $6.942 \times 10^{-10}$ m² | $4.3 \times 10^{-5}$ | 0.851 |
| **Gyroid** | 0.54 | $9.711 \times 10^{-10}$ m² | $6.1 \times 10^{-5}$ | 0.961 |

**Table 9: Maximum velocity and pressure gradient for different Reynolds number for Type 3 treatment of solid region.**

| | Diamond | | I-WP | | Primitive | | Gyroid | |
|---|---|---|---|---|---|---|---|---|
| $\dot{m}$ (kg/s) | Max. velocity (m/s) | Pressure gradient (kPa/m) | Max. velocity (m/s) | Pressure gradient (kPa/m) | Max. velocity (m/s) | Pressure gradient (kPa/m) | Max. velocity (m/s) | Pressure gradient (kPa/m) |
| **4e-8** | 2.3e-05 | -4.6e-03 | 4.3e-05 | -5.9e-03 | 5.8e-05 | -3.2e-03 | 2.3e-05 | -2.9e-03 |
| **4e-7** | 2.3e-04 | -4.6e-02 | 4.3e-04 | -5.9e-02 | 5.8e-04 | -3.2e-02 | 2.3e-04 | -2.9e-02 |
| **4e-6** | 2.3e-03 | -4.6e-01 | 4.3e-03 | -5.8e-01 | 5.8e-03 | -3.3e-01 | 2.3e-03 | -2.9e-01 |
| **4e-5** | 2.3e-02 | -4.6 | 4.3e-02 | -6.0 | 5.6e-02 | -3.4 | 2.3e-02 | -3.1 |
| **4e-4** | 2.2e-01 | -54.9 | 3.8e-01 | -72.7 | 4.6e-01 | -56.2 | 2.2e-01 | -45.4 |

The pressure drop with different orders of mass flow rate for Type 3 treatment of solid region is shown in Fig. 17. Pressure drop is observed to be maximum and minimum for the I-WP and Gyroid lattices respectively for the entire range of Reynolds number. Pressure drop in case of Primitive lattice is same as the Gyroid lattice for the low flow rate, $\dot{m} < 4e - 05 \, kg/s$ (or $Re < 10$), but sharply increases to match with the I-WP lattice during high flow rate $\dot{m} = 4e - 04 \, kg/s$ ($Re = 100$). This trend can also be explained using the large values of the deviation bars as shown in the Fig. 16 in which the deviation between pressure drop and linear Darcy law sharply increases for the Primitive case which leads to higher inertial resistance.

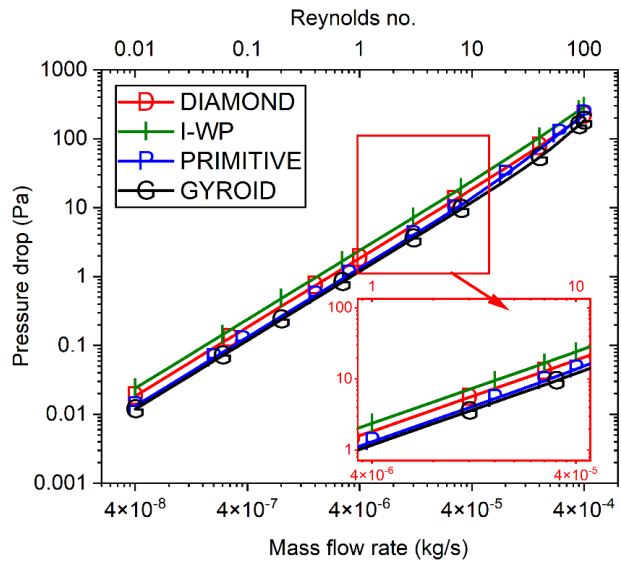

**Figure 17: Variation of pressure drop with Reynolds number for lattices with Type 3 scenario.**

*5.4 Comparison between three scenarios of solid region*

The I-WP Type 1 lattice has the highest pressure drop and thus the lowest flow conductance of all 12 cases as seen by its lowest permeability value (refer to Fig. 18 (a and b)). Primitive Type 2 has the lowest pressure drop and the highest flow conductance. Due to the increased fraction of volume available for the flow, Type 2 lattices have a 1-order lower pressure drop and a 1-order higher permeability. Furthermore, due to the increased cross-sectional area available for the flow, the maximum velocity for Type 2 lattice is reduced by one order compared to Type 1 and Type 3 lattices. When the inertial drag factors for all scenarios are evaluated, Gyroid Type 1 comes out on top with a value of 1.173. The inertial drag factor follows the same pattern in Type 1 and Type 3 lattices, with Gyroid being the biggest and I-WP being the smallest values among their types. The inertial drag factor for Type 2 lattices is much lower than for Type 1 and Type 3 lattices. This could be owing to the lower flow rates investigated in this study, where viscous drag is more important than inertial or form drag. As 100 percent volume is theoretically accessible for flow, it comparatively resulted in low velocity through the two different zones in Type 2 lattices, the form drag is less pronounced than the viscous drag loss in pressure drops. Primitive Type 2, with a value of 0.074, has the lowest value of the pressure drops. However, on comparing all the results, it is observed that the pressure drop with mass flow rate linearly increases at a slow pace up to $\dot{m} = 4e - 05 \, kg/s$ or $Re = 10$, and thereafter it increases steeply in a non-linear fashion.

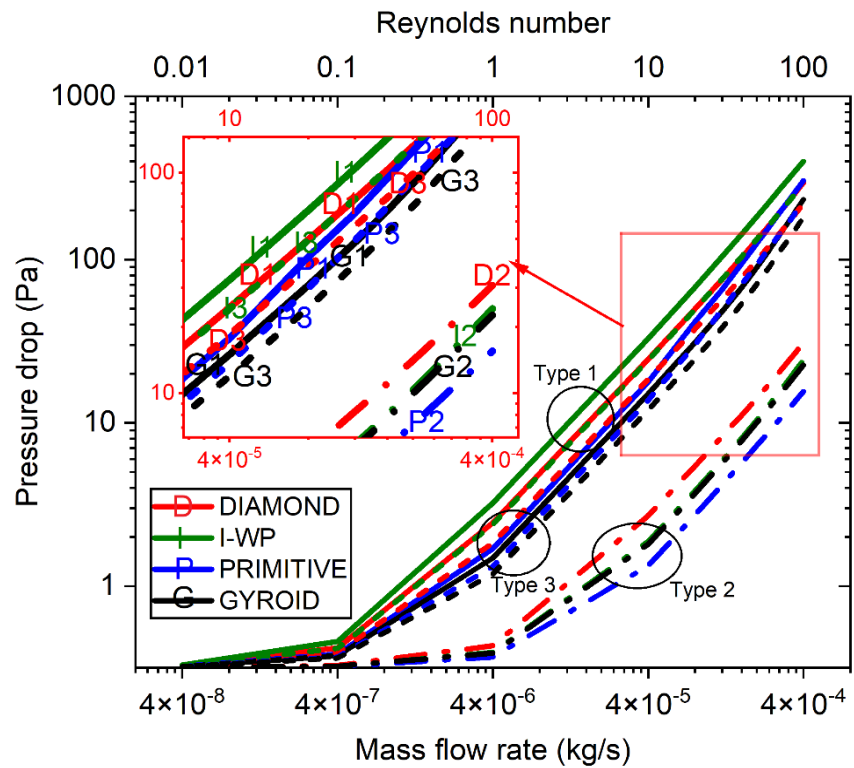

(a)

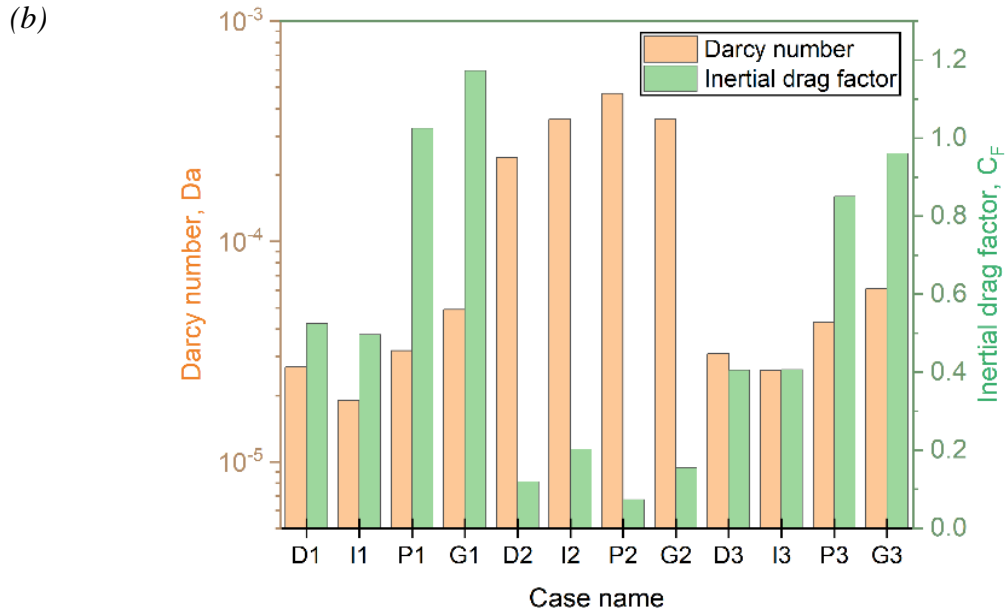

**Figure 18: (a) Variation of pressure drop with mass flow rate and Reynolds number, and (b) Darcy number and inertial drag factor for lattices with all three types of treatment for the solid region.**

For Type 1 porous media, the porosity is 0.32 and permeability is ranging from $3.045e-10$ to $7.892e-10\ m^2$, for which corresponding Darcy number is in the range of $1.9e-05$ to $4.9e-05$. Similarly, for Type 2 porous media in which the porosity is almost 1, however the volume contains zero thickness surfaces which offers the resistance to the flow, therefore the permeability is ranging from $3.898e-09$ to $7.505e-09\ m^2$ and corresponding Darcy number value is ranging from $2.4e-04$ to $4.7e-04$. Lastly, porosity is 0.54 for Type 3 porous media in which the permeability was ranging between $4.175e-10$ to $9.711e-10\ m^2$, for which corresponding Darcy number is in the range $2.6e-05$ to $6.1e-05$.

## 6. Conclusions

The flow within porous media and associated properties including permeability, drags (viscous and inertial), and pressure gradient, are presented by pore-scale numerical simulation for the TPMS lattices as the porous structures. For this work, four different types of TPMS lattices: Diamond, I-WP, Primitive, and Gyroid; and three different scenarios of the solid region of the lattice structure are considered. The solid region of the lattice is computationally treated as solid zone (Type 1), fluid zone (Type 2), and porous zone (Type 3), yielding a total of 12 different types of porous media, each of which is studied for hydrodynamic behavior for range of mass flow rates. Fig. 19 depicts the effect of effective porosity on the Darcy number and inertial drag factor. It is observed that the Darcy number (or dimensionless permeability) increases with the porosity, whereas the inertial drag factor follows the opposite trend and it decreases with the increase in porosity.

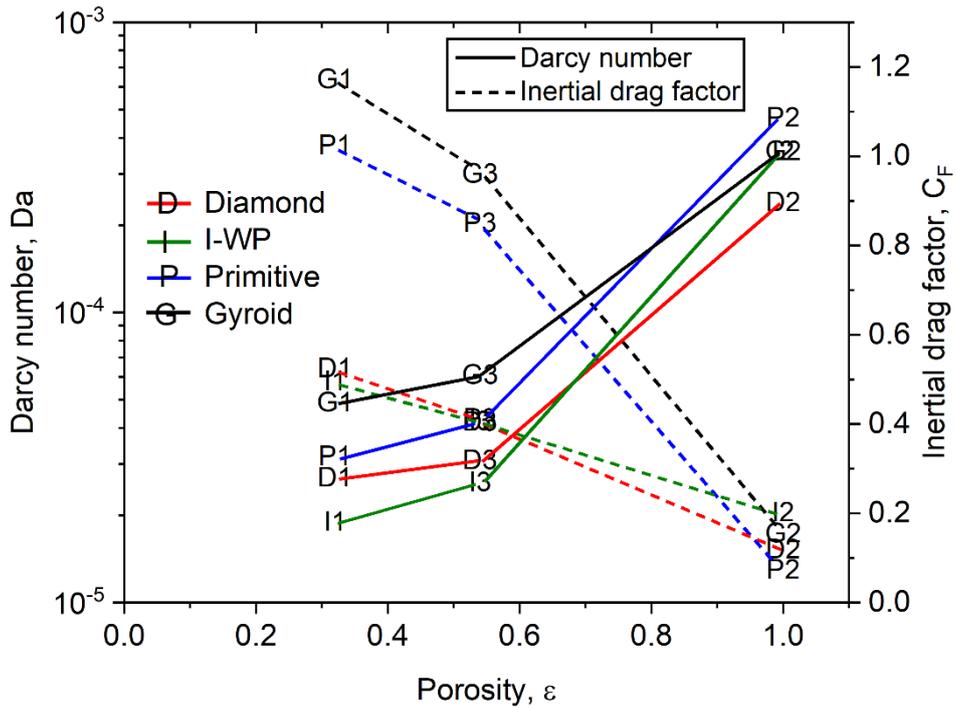

**Figure 19: Darcy number and inertial drag factor for lattices with all three types of treatment for the solid region.**

The key findings and conclusions from the present study are:

1. The Darcy flow or linear regime is only observed between a minimum $Re = 4$ (for Primitive Type 1 and 3) and maximum $Re = 20$ (for Diamond Type 2 and 3 lattice).
2. The maximum and minimum value of permeabilities is observed in Primitive Type 2 and I-WP Type 1 lattice respectively. In general, Type 2 lattices shows highest value of permeability and lowest value of inertial drag factor leading to lowest drop in pressure.
3. The Diamond, Primitive and Gyroid lattices have significant tortuosity in the flow channel for all three types of treatments for their solid zone, whereas the I-WP lattice is comparatively less tortuous.
4. The maximum pore velocity at mass flow rate $\dot{m} = 4e-04\ kg/s$ or corresponding $Re = 100$, is observed in the Primitive Type 1 lattice as $5.4e-01\ m/s$, whereas for the same flow rate, the minimum pore velocity is observed in Diamond Type 2 lattice as $7.5e-02\ m/s$.
5. Inertial drag factor is minimum for I-WP lattice at lower porosity and maximum for the Gyroid lattice, however, Primitive lattice offers minimum and I-WP lattice maximum at higher porosity.

# Nomenclature

| | | |
|---|---|---|
| $a$ | Unit cell size in x direction | [m] |
| $a_{sf}$ | Specific interfacial area | [m$^{-1}$] |
| $A$ | Cross-section area ($A = L^2$) | [m$^2$] |
| $b$ | Unit cell size in y direction | [m] |
| $c$ | Unit cell size in z direction | [m] |
| $C$ | Level-set constant | -- |
| $C_F$ | Inertial drag factor | -- |
| $d_p$ | Pore or particle size | [m] |
| $Da$ | Darcy number ($Da = K/L^2$) | -- |
| $K$ | Permeability | [m$^2$] |
| $L$ | Channel width | [m] |
| $\dot{m}$ | Mass flow rate | [kg/s] |
| $n$ | Normal distance from the surface | [m] |
| $p$ | Pore level pressure | [Pa] |
| $P$ | Average pressure | [Pa] |
| $Re$ | Reynolds number ($\rho U L/\mu$) | -- |
| $R$-$square$ | Coefficient of determination | -- |
| $u$ | Pore level velocity in x direction | [m/s] |
| $v$ | Pore level velocity in y direction | [m/s] |
| $w$ | Pore level velocity in z direction | [m/s] |
| $U$ | Average velocity in x direction | [m/s] |
| $V$ | Average velocity in y direction | [m/s] |
| $V_f$ | Volume of fluid region | [m$^3$] |
| $V_s$ | Volume of solid region | [m$^3$] |
| $W$ | Average velocity in z direction | [m/s] |
| $x$ | x-direction distance | [m] |
| $y$ | y-direction distance | [m] |
| $z$ | z-direction distance | [m] |
| $X$ | Lattice size in x direction | [m] |
| $Y$ | Lattice size in y direction | [m] |
| $Z$ | Lattice size in z direction | [m] |
| **Greek letters** | | |
| $\varepsilon$ | Porosity | -- |
| $\varepsilon_o$ | Porosity of the lattice | -- |
| $\varepsilon^*$ | Micro-porosity | -- |
| $\varepsilon_{eff}$ | Effective porosity of the lattice | |
| $\mu$ | Dynamic viscosity of the fluid | [kg/m-sec] |
| $\mu^*$ | Effective viscosity of porous medium | [kg/m-sec] |
| $\rho$ | Density of the fluid | [kg/m$^3$] |
| $\tau$ | Tortuosity | -- |
| **Subscripts** | | |
| solid | Property of solid region in the lattice | |
| void | Property of void region in the lattice | |

# References

[1] Stephen Whitaker, "Diffusion and Dispersion in Porous Media," *AIChE J.*, 1967.

[2] K. Vafai, "Analysis of the channeling effect in variable porosity media," *J. Energy Resour. Technol. Trans. ASME*, vol. 108, no. 2, pp. 131–139, 1986, doi: 10.1115/1.3231252.

[3] R. G. Carbonell and S. Whitaker, "Heat and mass transfer in porous media," in *Fundamentals of transport phenomena in porous media*, Springer, 1984, pp. 121–198.

[4] M. Kaviany, *Principles of heat transfer in porous media*. Springer Science & Business Media, 2012.

[5] K. VAFAI and A. AMIRI, *Non-Darcian Effects in Confined Forced Convective Flows*. Elsevier Science Ltd, 1998.

[6] S. Kandlikar, S. Garimella, D. Li, S. Colin, and M. R. King, *Heat transfer and fluid flow in minichannels and microchannels*. elsevier, 2005.

[7] A. Hadim, "Forced convection in a porous channel with localized heat sources," 1994.

[8] J. C. Ward, "Turbulent flow in porous media," *J. Hydraul. Div.*, vol. 90, no. 5, pp. 1–12, 1964.

[9] K. Hooman and M. Gorji-Bandpy, "Laminar dissipative flow in a porous channel bounded by isothermal parallel plates," *Appl. Math. Mech. (English Ed.*, vol. 26, no. 5, pp. 587–593, 2005, doi: 10.1007/BF02466332.

[10] Y. F. Maydanik, "Loop heat pipes," *Appl. Therm. Eng.*, vol. 25, no. 5, pp. 635–657, 2005, doi: https://doi.org/10.1016/j.applthermaleng.2004.07.010.

[11] S. Launay, V. Sartre, and J. Bonjour, "Parametric analysis of loop heat pipe operation: a literature review," *Int. J. Therm. Sci.*, vol. 46, no. 7, pp. 621–636, 2007, doi: 10.1016/j.ijthermalsci.2006.11.007.

[12] M. S. Ali, N. Pandey, M. Hadj-Nacer, M. Greiner, and M. F. Riyad, "Parametric study of two-phase flow in a porous wick of a mechanically pumped loop heat pipe," *AIP Conf. Proc.*, vol. 2324, no. February, 2021, doi: 10.1063/5.0037715.

[13] D. Butler, J. Ku, and T. Swanson, "Loop heat pipes and capillary pumped loops-an applications perspective," pp. 49–56, 2007, doi: 10.1063/1.1449707.

[14] K. C. Leong, C. Y. Liu, and G. Q. Lu, "Characterization of Sintered Copper Wicks Used in Heat Pipes," *J. Porous Mater.*, vol. 4, no. 4, pp. 303–308, 1997, doi: 10.1023/A:1009685508557.

[15] D. Deng, D. Liang, Y. Tang, J. Peng, X. Han, and M. Pan, "Evaluation of capillary performance of sintered porous wicks for loop heat pipe," *Exp. Therm. Fluid Sci.*, vol. 50, pp. 1–9, 2013, doi: 10.1016/j.expthermflusci.2013.04.014.

[16] S. Ergun and A. A. Orning, "Fluid flow through randomly packed columns and fluidized beds," *Ind. Eng. Chem.*, vol. 41, no. 6, pp. 1179–1184, 1949.

[17] D. A. Nield and A. Bejan, *Convection in porous media*, vol. 3. Springer, 2006.

[18] R. C. Givler and S. A. Altobelli, "A determination of the effective viscosity for the Brinkman-Forchheimer flow model," *J. Fluid Mech.*, vol. 258, no. 1994, pp. 355–370,

1994, doi: 10.1017/S0022112094003368.

[19] S. Liu, A. Afacan, and J. Masliyah, "Steady incompressible laminar flow in porous media," *Chem. Eng. Sci.*, vol. 49, no. 21, pp. 3565–3586, 1994, doi: 10.1016/0009-2509(94)00168-5.

[20] W. P. Breugem, "The effective viscosity of a channel-type porous medium," *Phys. Fluids*, vol. 19, no. 10, 2007, doi: 10.1063/1.2792323.

[21] J. Hommel, E. Coltman, and H. Class, "Porosity–Permeability Relations for Evolving Pore Space: A Review with a Focus on (Bio-)geochemically Altered Porous Media," *Transp. Porous Media*, vol. 124, no. 2, pp. 589–629, 2018, doi: 10.1007/s11242-018-1086-2.

[22] S. Whitaker, "Flow in porous media II: The governing equations for immiscible, two-phase flow," *Transp. Porous Media*, vol. 1, no. 2, pp. 105–125, 1986, doi: 10.1007/BF00714688.

[23] P. H. Valvatne, M. Piri, X. Lopez, and M. J. Blunt, "Predictive pore-scale modeling of single and multiphase flow," *Transp. Porous Media*, vol. 58, no. 1–2, pp. 23–41, 2005, doi: 10.1007/s11242-004-5468-2.

[24] C. Y. Wang, "A fixed-grid numerical algorithm for two-phase flow and heat transfer in porous media," *Numer. Heat Transf. Part B Fundam.*, vol. 32, no. 1, pp. 85–105, 1997, doi: 10.1080/10407799708915000.

[25] M. J. Blunt *et al.*, "Pore-scale imaging and modelling," *Adv. Water Resour.*, vol. 51, pp. 197–216, 2013, doi: 10.1016/j.advwatres.2012.03.003.

[26] R. A. Dawe, E. G. Mahers, and J. K. Williams, "Pore Scale Physical Modeling of Transport Phenomena in Porous Media," *Adv. Transp. Phenom. Porous Media*, pp. 47–76, 1987, doi: 10.1007/978-94-009-3625-6_3.

[27] V. Cromwell, D. J. Kortum, and D. J. Bradley, "The use of a medical computer tomography (CT) system to observe multiphase flow in porous media," *Proc. - SPE Annu. Tech. Conf. Exhib.*, vol. 1984-Septe, no. 1, pp. 1–4, 1984, doi: 10.2523/13098-ms.

[28] S. Y. Wang, S. Ayral, and C. C. Gryte, "Computer-Assisted Tomography for the Observation of Oil Displacement in Porous Media.," *Soc. Pet. Eng. J.*, vol. 24, no. 1, pp. 53–55, 1984, doi: 10.2118/11758-PA.

[29] H. Dong and M. J. Blunt, "Pore-network extraction from micro-computerized-tomography images," *Phys. Rev. E - Stat. Nonlinear, Soft Matter Phys.*, vol. 80, no. 3, pp. 1–11, 2009, doi: 10.1103/PhysRevE.80.036307.

[30] C. J. Elkins and M. T. Alley, "Magnetic resonance velocimetry: Applications of magnetic resonance imaging in the measurement of fluid motion," *Exp. Fluids*, vol. 43, no. 6, pp. 823–858, 2007, doi: 10.1007/s00348-007-0383-2.

[31] M. C. Gao, D. B. Miracle, D. Maurice, X. Yan, Y. Zhang, and J. A. Hawk, "High-entropy functional materials," *J. Mater. Res.*, vol. 33, no. 19, pp. 3138–3155, Oct. 2018, doi: 10.1557/jmr.2018.323.

[32] R. J. Hill, D. L. Koch, and A. J. C. Ladd, "Moderate-Reynolds-number flows in ordered and random arrays of spheres," *J. Fluid Mech.*, vol. 448, pp. 243–278, 2001, doi: 10.1017/s0022112001005936.

[33]  W. Chen, C. Ji, M. M. Alam, J. Williams, and D. Xu, "Numerical simulations of flow past three circular cylinders in equilateral-triangular arrangements," *J. Fluid Mech.*, pp. 1–44, 2020, doi: 10.1017/jfm.2020.124.

[34]  R. Ranjan, A. Patel, S. V. Garimella, and J. Y. Murthy, "Wicking and thermal characteristics of micropillared structures for use in passive heat spreaders," *Int. J. Heat Mass Transf.*, vol. 55, no. 4, pp. 586–596, 2012, doi: 10.1016/j.ijheatmasstransfer.2011.10.053.

[35]  X. Yang *et al.*, "Direct numerical simulation of pore-scale flow in a bead pack: Comparison with magnetic resonance imaging observations," *Adv. Water Resour.*, vol. 54, pp. 228–241, 2013, doi: 10.1016/j.advwatres.2013.01.009.

[36]  X. Yang *et al.*, "Intercomparison of 3D pore-scale flow and solute transport simulation methods," *Adv. Water Resour.*, vol. 95, pp. 176–189, 2016, doi: 10.1016/j.advwatres.2015.09.015.

[37]  K. K. Bodla, J. Y. Murthy, and S. V. Garimella, "Microtomography-based simulation of transport through open-cell metal foams," *Numer. Heat Transf. Part A Appl.*, vol. 58, no. 7, pp. 527–544, 2010, doi: 10.1080/10407782.2010.511987.

[38]  S. Krishnan, J. Y. Murthy, and S. V. Garimella, "A two-temperature model for solid-liquid phase change in metal foams," *J. Heat Transfer*, vol. 127, no. 9, pp. 995–1004, 2005, doi: 10.1115/1.2010494.

[39]  O. Al-Ketan and R. K. Abu Al-Rub, "Multifunctional mechanical metamaterials based on triply periodic minimal surface lattices," *Adv. Eng. Mater.*, vol. 21, no. 10, p. 1900524, 2019.

[40]  O. Al-Ketan and R. K. Abu Al-Rub, "MSLattice: A free software for generating uniform and graded lattices based on triply periodic minimal surfaces," *Mater. Des. Process. Commun.*, vol. 3, no. 6, pp. 1–10, 2021, doi: 10.1002/mdp2.205.

[41]  Y. Jung, K. T. Chu, and S. Torquato, "A variational level set approach for surface area minimization of triply-periodic surfaces," *J. Comput. Phys.*, vol. 223, no. 2, pp. 711–730, 2007.

[42]  O. Al-Ketan, R. K. Abu Al-Rub, and R. Rowshan, "The effect of architecture on the mechanical properties of cellular structures based on the IWP minimal surface," *J. Mater. Res.*, vol. 33, no. 3, pp. 343–359, 2018, doi: 10.1557/jmr.2018.1.

[43]  G. Hetsroni, M. Gurevich, and R. Rozenblit, "Sintered porous medium heat sink for cooling of high-power mini-devices," *Int. J. Heat Fluid Flow*, vol. 27, no. 2, pp. 259–266, 2006, doi: 10.1016/j.ijheatfluidflow.2005.08.005.

[44]  S. V. Patankar, C. H. Liu, and E. M. Sparrow, "Fully developed flow and heat transfer in ducts having streamwise-periodic variations of cross-sectional area," *J. Heat Transfer*, vol. 99, no. 2, pp. 180–186, 1977, doi: 10.1115/1.3450666.

[45]  ANSYS Fluent Tutorial Guide 18, "ANSYS Fluent Tutorial Guide 18," *ANSYS Fluent Tutor. Guid. 18*, vol. 15317, no. April, pp. 724–746, 2018.

[46]  M. Bahrami, M. M. Yovanovich, and J. R. Culham, "ICMM2005-75109," 2005.

[47]  R. K. Shah, "A correlation for laminar hydrodynamic entry length solutions for circular and noncircular ducts," 1978.

[48] W. S. Almalki and M. H. Hamdan, "Investigations in effective viscosity of fluid in a porous medium," *J. Eng. Res. Appl. www.ijera.com*, vol. 6, no. April, pp. 41–51, 2016, [Online]. Available: www.ijera.com.

[49] B. Alazmi and K. Vafai, "Analysis of variants within the porous media transport models," *J. Heat Transfer*, vol. 122, no. 2, pp. 303–326, 2000, doi: 10.1115/1.521468.

# Appendix

## 1. *Equation for momentum conservation in porous media*

a. Darcy model (used for low flow rates, $Re_p < 1$)

$$\nabla p = -\frac{\mu}{K}\vec{V} \tag{A1}$$

b. Darcy Forchheimer model (used for low flow rates, $Re_p > 10$)

$$\nabla p = -\frac{\mu}{K}\vec{V} - \frac{\rho C_F}{\sqrt{K}}\vec{V}|\vec{V}| \tag{A2}$$

c. Darcy Forchheimer Brinkman model (used for high porosity, $\varepsilon \to 1$)

$$\nabla p = -\frac{\mu}{K}\vec{V} - \frac{\rho C_F}{\sqrt{K}}\vec{V}|\vec{V}| + \mu^* \nabla^2 \vec{V} \tag{A3}$$

In the Eq. (A3), $\mu^*$ is the effective viscosity of the porous medium and some important relations between the porosity and effective viscosity are mentioned in Table A1.

## 2. *Tables*

**Table A1: Different correlations for effective viscosity**[48].

| Sr. No. | Name of the correlation | Effective viscosity |
|---|---|---|
| 1. | Einstein's relation | $\mu^* = \mu\left(1+\frac{5}{2}(1-\varepsilon)\right)$ |
| 2. | Brinkman's relation | $\mu^* = \mu\left(\frac{1}{\varepsilon}\right)$ |
| 3. | Breugem's relation | $\mu^* = \mu\left(\frac{1}{2}\left(\varepsilon-\frac{3}{7}\right)\right)$ |
| 4. | Common assumption in high porosity media | $\mu^* = \mu$ |

**Table A2: Different formulations for Darcy-Forchheimer-Brinkman model**[49].

| Sr. No. | Model version | Darcy term | Forchheimer term | Brinkman term |
|---|---|---|---|---|
| 1. | V1 | $-\frac{\mu}{K}\vec{V}$ | $-\frac{\rho C_F}{\sqrt{K}}|\vec{V}|\vec{V}$ | $\mu \nabla^2 \vec{V}$ |
| 2. | V2 | $-\frac{\mu}{K}\vec{V}$ | $-\frac{\rho C_F \varepsilon}{\sqrt{K}}|\vec{V}|\vec{V}$ | $\frac{\mu}{\varepsilon}\nabla^2\vec{V}$ |
| 3. | V3 | $-\frac{\mu}{K}\vec{V}$ | $-\frac{\rho C_F}{\sqrt{K}}|\vec{V}|\vec{V}$ | $\frac{\mu}{\varepsilon}\nabla^2\vec{V}$ |

**Table A3: Properties of porous medium used in Type 3 treatment**[43]**.**

| Property | Value |
|---|---|
| Porosity, $\varepsilon$ | 0.32 |
| Permeability, K | $0.215 \times 10^{-10}$ m$^2$ |
| Inertial drag factor, $C_F$ | 1.23 |

### 3. Calculation of tortuosity

Tortuosity is calculated by taking two points in the direction of the flow, and measuring actual length of flow path (length of fluid trajectory) and minimum length (length of straight line) between these two points. It is then defined as below:

$$\tau = \frac{length\ of\ actual\ path}{length\ of\ minimum\ path} \tag{A4}$$

### 4. Calculations of the slope and curvature of the curve

Using the Taylor series expansion, any polynomial is approximated up to 2$^{nd}$ order term as:

$$f(x) = f(0) + \frac{f'(0)}{1!}x + \frac{f''(0)}{2!}x^2$$

In this expression, first term (or zeroth order derivative) indicates the intercept of the curve and the coefficient of second and third terms calculates the slope (first order derivative) and curvature (second order derivative) of the curve respectively.

In addition to that, the Darcy-Forchheimer law is a second order function between the pressure gradient and the mass flow rate. Since the pressure drop is zero in case of no flow, the intercept is also zero in the resulting curve.

$$\frac{\Delta P}{L} = -\left(\frac{\mu}{\rho A K}\right)\dot{m} - \left(\frac{C_F}{\rho A^2 \sqrt{K}}\right)\dot{m}^2$$

On comparing the coefficients of first and second order terms, it can be concluded that the slope of the curve is inversely proportional to the permeability and the curvature of the curve is directly proportional to the inertial drag factor of the porous medium.